\documentclass[preprint,11pt]{elsarticle}
\usepackage[a4paper, top=1in, bottom=1in, left=1in, right=1in]{geometry}
 
\makeatletter
\def\ps@pprintTitle{%
 \let\@oddhead\@empty
 \let\@evenhead\@empty
 \def\@oddfoot{\centerline{\thepage}}%
 \let\@evenfoot\@oddfoot}
\makeatother
\usepackage{amssymb,amsmath}
\usepackage{hyperref}

\usepackage{caption}
\usepackage{float}
\usepackage{booktabs}
\usepackage{adjustbox}
\usepackage{longtable}
\usepackage{indentfirst} 
\usepackage{soul}
\usepackage[titletoc]{appendix}
\usepackage{xcolor}
\usepackage[T1]{fontenc}
\usepackage{comment}
\usepackage{float} % Enable the position of figures
\usepackage{placeins}  % Enable the position of figures

\makeatletter
\renewcommand{\corref}[1]{\textsuperscript{*}}
\makeatother

\usepackage{ulem}
\begin{document}
\captionsetup[figure]{labelfont={bf},labelformat={default},labelsep=period,name={Fig.}}

\begin{frontmatter}

 \title{Micromechanics of Nanoporous Alloying Electrodes}

%Referees: Aashutosh Mistry, Bai-Xiang Xu, Wendy Gu
% Experimental referee: Eric Detsi, Jian Luo
% Delin's work too 

\author[mymainaddress]{Yu-Cheng Lai\textsuperscript{+}}
\author[USC]{Delin Zhang\textsuperscript{+}}
\author[UCLA]{Kodi Thurber}
\author[UCLA]{Kai Smith}
\author[SLAC]{Johanna Nelson Weker}
\author[UCLA]{Sarah Tolbert}
\author[mymainaddress]{Ananya Renuka Balakrishna\corref{cor1}}
\cortext[cor1]{Corresponding author.}
\ead{ananyarb@ucsb.edu}
\address[mymainaddress]{Materials Department, University of California Santa Barbara, USA}
\address[USC]{Aerospace and Mechanical Engineering, University of Southern California, USA}
\address[UCLA]{Department of Chemistry and Biochemistry, University of California Los Angeles, USA}
\address[SLAC]{Stanford Synchrotron Radiation Lightsource, SLAC National Accelerator Laboratory, USA}

\vspace{-5mm}
\begin{abstract}
\small{Nanostructured electrodes with voids or interconnected pores accommodate large volume changes, shorten ion diffusion pathways, and enhance the structural reversibility of alloying electrodes. While these nanoporous features improve the performance of architected electrodes over bulk electrodes, they also act as geometric irregularities that localize and concentrate internal stresses. In this work, we investigate the hierarchical interplay between phase boundaries and nanoporous features at the microstructural scale and their collective role in mitigating chemo-mechanical failure at the engineering scale. Using Sb$\to$Li$_2$Sb$\to$Li$_3$Sb as a model system, we develop a continuum framework coupling lithium diffusion and reaction kinetics with the finite deformation of alloying electrodes. We analytically show that the interfacial stresses (and not volume changes alone) induce fracture during the Sb $\to$ Li$_2$Sb transformation. Nanoporous geometries accommodate volume changes and in turn lower these interfacial stresses that are detrimental to the structural stability of architected electrodes. We analytically study the elastic-plastic deformation accompanying the Sb $\to$ Li$_2$Sb transformation, and develop a micromechanical model using a hyperelastic neo-Hookean material law to predict the deformations accompanying the Li$_2$Sb$\to$Li$_3$Sb transformation. Our results reveal how diffusion and reaction kinetics shape phase boundary morphology, predict stress concentrations at defects, identify crack geometries likely to propagate, and show how carefully architected electrodes relieve stresses. These findings highlight critical design principles to optimize electrode lifespan and demonstrate a potential application of our continuum model as an electrode design tool.}
\end{abstract}


\end{frontmatter}

\renewcommand{\thefootnote}{+}
\footnotetext{Both authors contributed equally to this work.}

\newpage
\section{Introduction}

%\noindent \textcolor{red}{incorporate references by Mistry group, Wendy Gu, Bai-Xiang Xu, and also some of Delin's work ;)}
\noindent One of the primary constraints on the widespread use of lithium-ion batteries is the availability of electrodes that offer both high-energy storage and long cycle life. Alloying-type electrodes, such as silicon, tin, and antimony, offer high specific capacities (theoretically 660 - 4200 mAh/g) \cite{zhang2011review}; however, they also undergo large and inhomogeneous volume changes during alloying reactions \cite{chao2011study,mcdowell201325th}. These structural changes generate significant interfacial stress and hydrostatic pressure, which, in turn, induces microcracking and electrode fragmentation \cite{shao2024resolving, cook2017nanoporous}. With repeated cycling, these changes lead to the loss of contact both within and between battery materials, accelerate side reactions, and contribute to eventual capacity loss in batteries.

\vspace{2mm}
\noindent Despite the chemo-mechanical challenges, alloying-type electrodes present a promising route as high-capacity anodes. Antimony (Sb), for example, offers high capacity and good rate capability with both Li$^+$ and Na$^+$ \cite{darwiche2012better}. When compared to alloying anodes, such as silicon (Si) or tin (Sn), Sb on lithiation undergoes relatively smaller volume changes ($\sim 125\%$) and capacity losses \cite{zhang2011review}. For example, Si-based and Sn-based anodes experience substantial volume changes (320$\%$ and 260$\%$, respectively) from the pristine phase to the fully-lithiated phase \cite{zhang2011review}, which are nearly 2--3 times larger than Sb. This often leads to the pulverization of active alloy particles and, consequently, poor cycling stability. Furthermore, antimony operates at moderate voltages 0.9~V vs. Li$^+$/Li and is readily alloyable with other metals, such as the Bi alloy \cite{zhao2015high} or Sn alloy \cite{lin2020understanding, xie2018beta}. These binary alloys, in turn, have favorable electrochemical properties including a flat voltage curve and stable cycling performance. 

\vspace{2mm}
\noindent Besides favorable electrochemical performance, antimony undergoes fewer phase transformations when reacting with Li$^+$, making it a good candidate for a theoretical study. Inserting Li$^+$ into Sb is a simple two-step process, in which a Li$_2$Sb phase forms at $0.87~ \mathrm{V}$, and with continued lithiation, Li$_3$Sb phase forms at $0.85~\mathrm{V}$ \cite{he2018antimony}. Both Li$_2$Sb and Li$_3$Sb are crystalline and the formation of only one intermediate phase (Li$_2$Sb), during the alloying reaction, contributes to improved cyclability of the electrode \cite{he2018antimony}. This alloying reaction is accompanied by two other physical processes, namely Li-diffusion in the pristine/reacted phases and finite deformation of the electrode. Li$^+$ diffuses through the electrode during charge/discharge cycles, and the competing kinetics of Li-diffusion and Li-Sb reaction define the morphology of the phase boundary \cite{afshar2021thermodynamically}. These diffusion and reaction mechanisms collectively determine phase transformation kinetics and mechanical deformation of the electrode. On Li$^+$ removal, the Li$_3$Sb phase transforms directly to Sb phase, and the intermediate Li$_2$Sb phase is not observed.\footnote{This asymmetry in the chemical reactions during lithiation and delithiation of Sb is attributed to its phase transformation hysteresis \cite{chang2015elucidating}. Sodiating Sb, however, generates multiple intermediate phases, some of which are amorphous in nature \cite{allan2016tracking}. These amorphous intermediates are theorized to accommodate internal stresses, contributing to the improved reversibility of Sb. In this study, however, we confine our investigation to large-volume changes in anodes in which all phases are crystalline.}

\vspace{2mm}
\noindent In addition to the coupled diffusion-reaction process, the two-step phase transformations in Sb are accompanied by finite deformation (i.e., large deformation) of the electrode. Inserting Li$^+$ into Sb induces a rhombohedral (Sb) to a hexagonal (Li$_2$Sb) structural transformation that is accompanied by $\sim90\%$ volume change \cite{chang2015elucidating}. Continued lithiation transforms the lattices to a cubic symmetry (Li$_3$Sb) and a further 24\% in volume change \cite{baggetto2013intrinsic}. These structural transformations not only generate enormous volume changes but significant lattice misfits between Sb/Li$_2$Sb and Li$_2$Sb/Li$_3$Sb phases. The lattice misfit strains, in turn, induce plastic yielding and accelerate microcracking and fracturing of Sb electrodes \cite{he2018antimony, corsi2021insights, chang2015elucidating}. These large volume changes and internal stresses are widely observed in several alloying and conversion types of anodes \cite{zhang2011review} and the resulting chemo-mechanical degradation contributes to side reactions, accelerates capacity fade, and shortens battery lifespans.

\vspace{2mm}
\noindent Researchers have shown that strategies such as nanostructuring and intermetallic alloying of electrodes mitigate chemo-mechanical degradation \cite{cook2017using, li2020impact, hou2015sb, lin2020understanding, xie2018beta, mistry2019deconstructing, mistry2019mesoscale}. Nanostructuring of Sb electrodes, such as designing porous architectures, accommodates large volume changes during alloying reactions and improves material reversibility \cite{cook2017using}. Hierarchical nanoporous architectures with bimodal porosity (i.e., containing two sizes of pores) not only show an improvement in structural reversibility but are also reported to have shorter diffusion distances that contribute to improved kinetics \cite{li2020impact, hou2015sb}. Intermetallic alloys typically undergo sequential phase transformations, which in turn facilitates the buffering of internal stresses and contributes to improved reversibility \cite{lin2020understanding, xie2018beta}. Other strategies, such as alloying with Na$^+$, are shown to generate intermediate phases (that have no long-range order), which alleviate internal stresses and enhance material lifespans \cite{allan2016tracking, gutierrez2019interpreting, selvaraj2020remarkable}. These strategies have greatly helped in improving lifespans of alloying electrodes, however, we do not clearly understand the microstructural deformation mechanisms that collectively mitigate chemo-mechanical failure. It is important to understand the stress fields that accompany phase transformation pathways and their interplay with defects, such as cracks and pores, during an alloying reaction. By doing so, we would gain fundamental insights into how large-volume changes and interfacial stresses evolve and accumulate at defects with repeated cycling, and thus guide the microstructural designing of battery materials.

\vspace{2mm}
\noindent Operando characterization of alloying reactions (e.g., transmission X-ray microscopy (TXM) \cite{lin2020understanding}, solid-state NMR \cite{allan2016tracking}, XRD \cite{cook2017using,lin2020understanding,nelson2012operando}) provides crucial insights into phase transformation pathways, strain distributions across multiple electrode particles, and microstructural failure mechanisms \cite{paul2021review, nelson2015emerging}.
These studies reveal that nanostructured electrodes offer better structural reversibility than bulk electrodes \cite{lin2020understanding}, and that the cycling life of electrodes is independent of pore geometries \cite{li2020impact}. 
Moreover, experimental measurements thus far have reported homogenized responses of nanostructured electrodes and often attribute failure to the large volume changes during electrochemical cycling. We do not fully understand the microstructural interplay between a moving phase boundary and defects. For example, we do not know how volume changes are accommodated into pores and whether and how interfacial stresses are relieved around cracks. This limits us from geometrically designing electrodes with suitable defect geometries and distributions that are necessary to minimize internal stresses.  

\vspace{2mm}
\noindent Recent works on optimizing electrode geometries show that architecting electrodes is not only important to enhance energy storage density \cite{li2024topology}, but are also important for reducing diffusion lengths and enabling structural reversibility during electrochemical cycling \cite{baggetto2011honeycomb,li2014mesoporous,yao2011interconnected}. For example, architecting electrodes in the form of honeycomb structures shows how \textcolor{black}{geometric instabilities (e.g., bending or snapping)} can be programmed to improve the structural reversibility of electrodes despite large deformations. At this engineering scale, it is not only important to understand how phase boundaries interact with defects such as cracks and pores but also to understand how \textcolor{black}{geometric instabilities (e.g., bending or snapping)} can be programmed into electrode structures to improve their structural reversibility.

\vspace{2mm}
\noindent 
Theoretical models developed for analogous battery materials, such as silicon and FeS$_2$, provide quantitative insights into the microstructural mechanisms and stress distributions leading to failure in these systems \cite{ afshar2021thermodynamically,zhao2010fracture,pandurangi2024chemo,salvadori2018coupled,svendsen2018finite, zhao2019diffusion,grossman2023topology,zhao2011concurrent,zuo2020rational, zhao2019review}. For example, the tensile hoop stress evolving on the surface of lithiated silicon is a key reason for microcracking in these electrodes \cite{zhao2011concurrent}. A similar interplay between phase boundaries and crack tips is shown in the case of intercalation compounds such as Li$_x$CoO$_2$ \cite{zhao2010fracture} and niobium tungsten oxide \cite{pandurangi2024chemo}. In the SiC/SiO$_2$ system \cite{zhao2019diffusion} and FeS$_2$ electrodes \cite{afshar2021thermodynamically}, a sharp reaction interface generates significant tensile stresses around the reaction front that are sufficient to propagate surface cracks. By contrast, features such as pores or cavities in nanoporous electrodes are designed to accommodate volume changes accompanying alloying reactions \cite{cook2017nanoporous, grossman2023topology, zuo2020rational}. Theoretical studies report the role of pore-sizes and wall-thickness in suppressing failure \cite{grossman2023topology, zuo2020rational}. However, these latter works \cite{grossman2023topology, zuo2020rational} do not distinguish ion diffusion from phase transformation kinetics, resulting in a loss of information regarding phase-boundary morphologies and interfacial stresses during electrochemical cycling \cite{afshar2021thermodynamically,salvadori2018coupled,svendsen2018finite,zhao2019diffusion}. Consequently, the delicate interplay between interfacial stresses and electrode architectures remains poorly explored. We build on previous modeling efforts not only to investigate the interaction between phase boundaries and defects within a reaction-diffusion-deformation framework, but also to understand how phase transformation pathways affect the structural reversibility of architected electrodes.

\vspace{2mm}
\noindent In this work, we combine electrochemical experiments and micromechanical theories to investigate the failure mechanisms accompanying alloying reactions. Specifically, we explore how microstructural phase transformation pathways, reaction and diffusion kinetics, and macroscopic electrode architectures collectively govern the structural reversibility of an alloying electrode. Using the Sb $\rightarrow$ Li$_2$Sb $\rightarrow$ Li$_3$Sb alloying reactions as representative examples, we investigate how electrode geometry affects microstructural evolution and performance. Specifically, we study how the nanoporous geometry of an Sb electrode interacts with evolving phase transformation microstructures and analyze the interfacial stresses that accompany inhomogeneous volume changes. Our continuum model, based on the \textcolor{black}{hyperelastic neo-Hookean law}, predicts the coupled interplay between reaction-diffusion kinetics and finite deformations, which govern microstructural phase transformation pathways in alloying anodes. Our results show that reaction and diffusion kinetics affect phase boundary morphologies, leading to critical stresses at crack tips. Additionally, we find that porous cavities help accommodate large volume changes and alleviate interfacial stresses during alloying. This interplay between microstructures and electrode geometry can be strategically engineered to mitigate the structural degradation of alloying anodes. Finally, we demonstrate a potential application of our continuum model as a design tool to engineer \textcolor{black}{geometric instabilities (e.g., bending)} in electrode architectures, accommodating large deformations with minimal internal stresses and thereby extending the electrode lifespan. 

\section{Experimental Investigation of the Alloying Reaction in Antimony}
\label{sec:experiments}
\noindent We first outline the experimental results on the synthesis, characterization, and electrochemical cycling of antimony electrodes with bulk and nanoporous geometries. A key finding is that nanoporous Sb (NP-Sb) electrodes retain 2.5 times more capacity than bulk Sb electrodes when cycled under identical conditions. These results also provide insights into nanoporous electrode architectures and their failure modes during lithiation. We use these measurements to calibrate our continuum model and compare our calculations to the experimental data in Section \ref{sec:results}.
\subsection{Synthesis, Electrode Preparation and Cell Assembly}
\noindent 
Bulk Sb samples (Fisher Scientific; antimony lump, 99.5$\%$ metals basis; 010095-22) were ground into a fine powder using a mortar and pestle and used as a control. NP-Sb particles were prepared by first synthesizing an Mg$_3$Sb$_2$ ingot. This ingot was prepared by mixing 1 g of elemental powders in a 3:2 Mg:Sb molar ratio in a graphite crucible and heating at 750 °C for 12 h under flowing argon. After cooling to room temperature, the ingot was ground using a mortar and pestle, and then immersed in 300 mL of a 3 wt$\%$ tartaric acid solution for 12 h to selectively dealloy Mg. The resulting dealloyed Sb powder was washed five times with DI water by gravity sedimentation, vacuum-dried for 6 h, and transferred into an argon-filled glovebox for storage.

\vspace{2mm}
\noindent 
Electrodes were prepared by casting a slurry containing 70 wt$\%$ active material, 15 wt$\%$ Super P conductive carbon, and 15 wt$\%$ binder (a 10 wt$\%$ aqueous solution of styrene–butadiene rubber (SBR) and carboxymethylcellulose (CMC) mixed at a 1:1 ratio) onto 9 $\mu$m copper foil using a doctor blade (25 $\mu$m wet-film thickness). The coated electrodes were dried under ambient conditions, calendered between two glass sheets, and then vacuum-dried at 120 °C for 12 h.

\vspace{2mm}
\noindent 
Electrochemical cells were assembled in two formats. Pouch cells used for operando X-ray diffraction (XRD) and Transmission X-ray Microscopy (TXM) were both assembled using aluminized Mylar packaging, 3 mm nickel tabs insulated with polymer tape, and 1 M LiPF$_6$ in ethylene carbonate/dimethyl carbonate(EC:DMC) with 10 vol$\%$ fluoroethylene carbonate (FEC) as the electrolyte. A Whatman GF/C glass-fiber separator was used for the operando XRD cells, whereas a thinner Celgard 2400 monolayer polypropylene separator was used for the TXM cells. The use of Celgard for TXM is to avoid out-of-focus features which interfere with the image quality. The CR2032 coin cells, used for galvanostatic cycling, were assembled using the prepared electrodes, stainless-steel spacers and springs, a Whatman GF/C glass-fiber separator, and 1 M LiPF$_6$ in EC:DMC with 10 vol$\%$ FEC as the electrolyte. All electrochemical measurements were performed in a half-cell configuration, with the Sb-based electrode serving as the working electrode and Li metal serving as the counter/reference electrode.

\subsection{Characterization and Electrochemical Cycling}
\begin{figure}[ht!]
    \centering
    \includegraphics[width=0.75\textwidth]{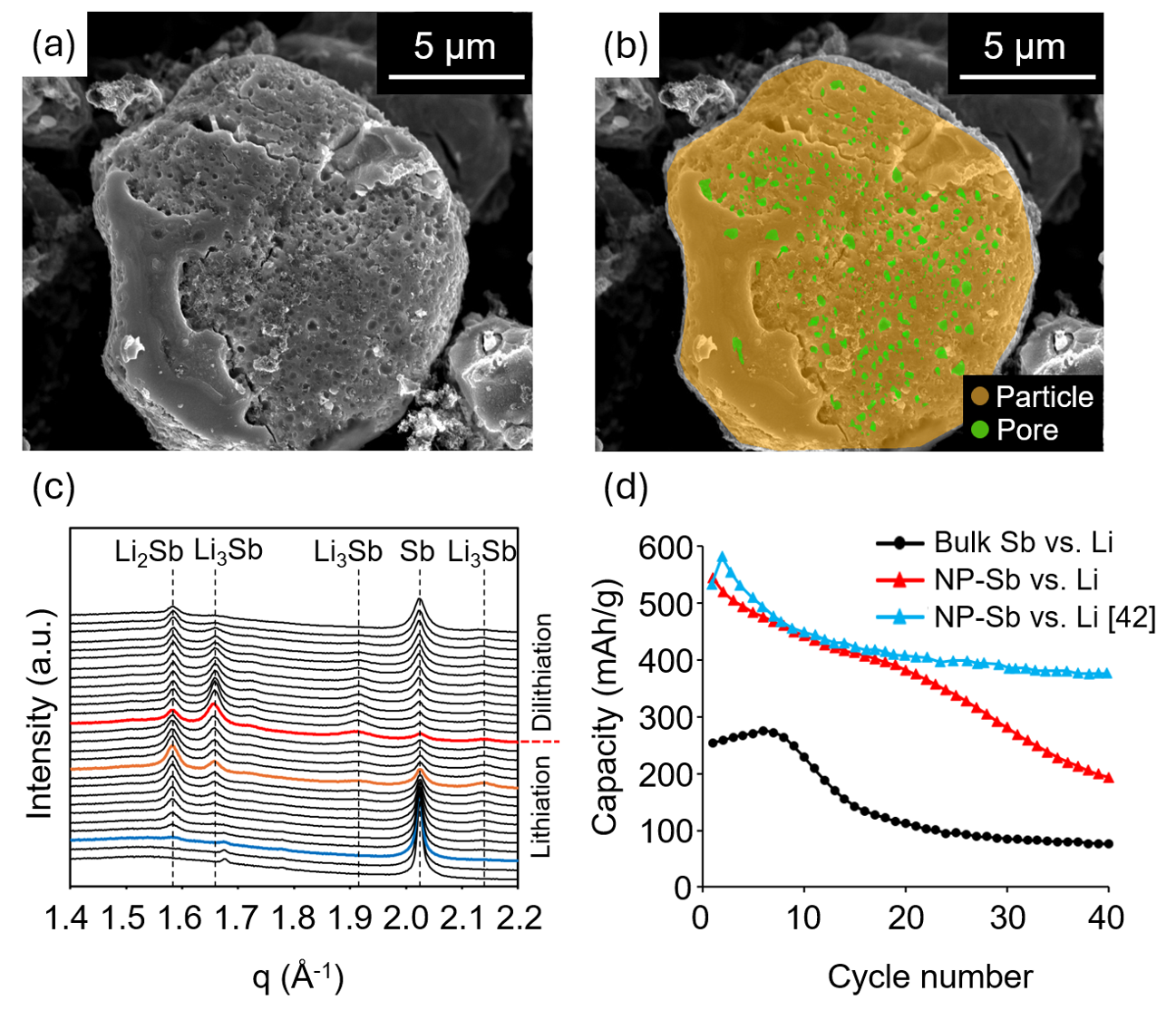}
    \caption{(a) Scanning electron microscopy (SEM) image of a representative nanoporous Sb (NP-Sb) particle. (b) Pore-size analysis based on the segmentation in (a), showing an average pore radius of 61.77 nm across the NP-Sb particle. (c) Operando X-ray diffraction (XRD) of an NP-Sb cell, showing the phase evolution Sb $\rightarrow$ Li$_2$Sb $\rightarrow$ Li$_3$Sb during cycling. The intensity measurements follow the first electrochemical cycle, beginning with lithiation from an open-circuit voltage to 0.05 V and followed by delithiation up to 1.5 V. The blue and orange traces, respectively, represent scanning at 0.82 V and 0.75 V. The red trace represents the fully lithiated state. On this red trace, multiple Li$_3$Sb reflections are observed, consistent with the coexistence of cubic and hexagonal polymorphs \cite{baggetto2013intrinsic,mayo2017structure}. (d) NP-Sb exhibits improved cycling stability compared with bulk Sb over 40 cycles \cite{thurber2025antimony}.
    \label{Exp_F1}}
\end{figure}

\noindent Fig.~\ref{Exp_F1}(a) shows a representative scanning electron microscopy (SEM) image of a NP-Sb particle. Contrast-limited adaptive histogram equalization (CLAHE) is applied to each particle to enhance the visibility of small dark features. The pore regions are then segmented using adaptive thresholding as shown in Fig.~\ref{Exp_F1}(b). The orange area represents the Sb particle and the green areas denote the segmented pores. For each pore, the equivalent 2D radius, $\sqrt{A/\pi}$, is projected onto its spherical surface. These projected radii are then used to compute pore geometries, yielding an average pore radius of 61.77 $\pm $ 18.14 nm.

\vspace{2mm}
\noindent
XRD measurements conducted at Stanford Synchrotron Radiation Lightsource (SSRL) Beamline 11-3 enabled real-time tracking of crystallographic phase evolution during electrochemical cycling.  Fig.~\ref{Exp_F1}(c) shows the XRD pattern of a NP-Sb electrode cell. The Sb reflection near $q\approx2.02~\mathring{\mathrm{A}}^{-1}$ progressively weakens, while a new reflection assigned to Li$_2$Sb emerges and grows at $q\approx1.60~\mathring{\mathrm{A}}^{-1}$, indicating a clear Sb $\to$ Li$_2$Sb transformation. Upon further lithiation, the Li$_2$Sb feature decreases and multiple reflections associated with Li$_3$Sb increase (at $q\approx1.66,1.90 \text{ and }2.14 ~\mathring{\mathrm{A}}^{-1}$), consistent with the commonly reported two-step alloying pathway Sb $\to$ Li$_2$Sb $\to$ Li$_3$Sb \cite{park2007high,hewitt2001electrochemistry}. Previous work has also noted that  Li$_3$Sb exhibits polymorphism (cubic and hexagonal) \cite{baggetto2013intrinsic,mayo2017structure}, resulting in multiple peaks for Li$_3$Sb.

\vspace{2mm}
\noindent 
Fig.~\ref{Exp_F1}(d) compares the cycling capacity of bulk Sb and NP-Sb electrodes at a C/5 rate between 0.05~V and 1.5~V versus Li/Li$^{+}$. The results show that NP-Sb delivers a higher specific capacity throughout cycling, starting from a high initial capacity and gradually fading to $\sim$193 mAh g$^{-1}$ after 40 cycles. In contrast, the bulk Sb electrode shows a much lower capacity, with a slight increase during the first 10 cycles followed by steady decay to 75.66 mAh g$^{-1}$ by 40 cycles. Recent work by Thurber et al. \cite{thurber2025antimony} reports an even more dramatic improvement in capacity for nanoporous antimony prepared by a slightly different synthesis route (Fig.~\ref{Exp_F1}(d) blue curve). Overall, NP-Sb maintains a clear capacity advantage over bulk Sb across the entire cycling window, indicating improved electrochemical performance.

\subsection{Operando 2D-Transmission X-Ray Microscopy}
\noindent We next probe phase transformation and microstructural evolution pathways in bulk and nano-porous antimony electrodes during electrochemical cycling. Specifically, we perform operando 2D-Transmission X-Ray Microscopy (TXM) imaging at the SSRL Beamline (BL) 6-2C to investigate the failure mechanisms in electrode single particles during lithiation. This technique enables users to image particles in functional pouch cell geometries, which are representative of standard battery geometries. Furthermore, this beamline provides a $30~\mu$m field of view with a pixel size of 30.4 nm directly below the Cu K edge, 8950 eV, which minimizes the absorption of the Cu foil. The microscope is suitable for observing crack growth and the eventual fracturing that accompanies the electrochemical cycling of antimony electrodes.
\begin{figure}[!ht]
    \centering
    \includegraphics[width=0.75\textwidth]{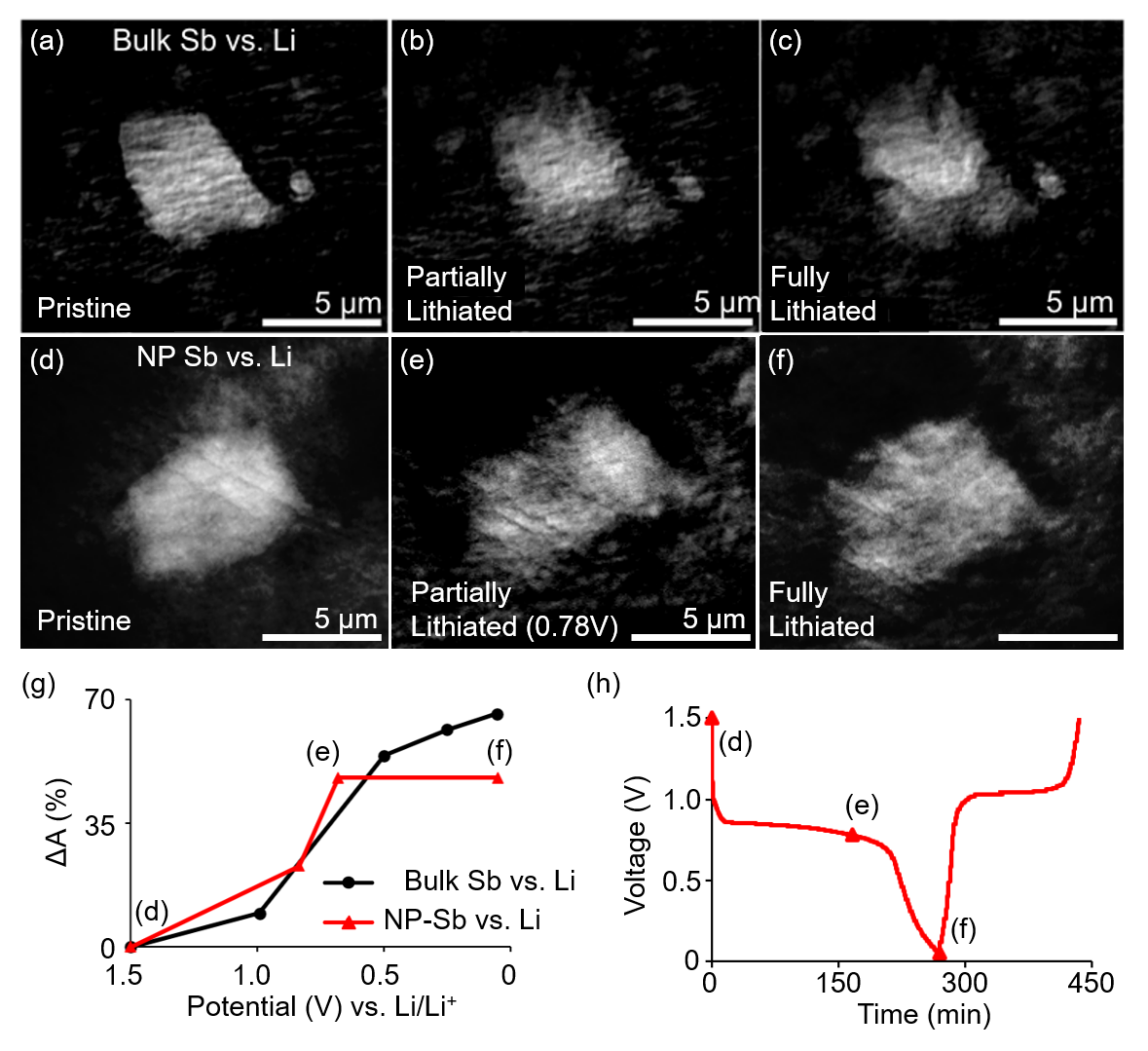}
    \caption{(a-c) TXM images showing the lithiation process of a bulk Sb particle, transforming from (a) pristine Sb to (b) partially lithiated phase and finally to (c) fully lithiated Li$_3$Sb. The measured average areal expansion $\Delta$A $\approx 66\%$ corresponds to an estimated volume expansion $\Delta$V $\approx$ $114\%$. This severe swelling fragments the bulk Sb particle into multiple pieces during lithiation. (d-f) In contrast, NP-Sb particles show a lower average areal expansion $\Delta$A $\approx$ $48\%$ and a reduced estimated volume expansion $\Delta$V $\approx$ 80$\%$, which results in fewer cracks in the NP-Sb electrode. (g) Projected area expansion data of bulk Sb (black line) and NP-Sb (red line) particles. Values for NP-Sb are averaged from six  particles. (h) Voltage–time profile during C/5 lithiation–delithiation of a half-cell using an NP-Sb anode, with the states corresponding to panels (d–f) labeled on the curve. }
    \label{Exp_F2}
\end{figure}

\vspace{2mm}
\noindent Fig.~\ref{Exp_F2}(a-c) shows the lithiation of a single bulk Sb particle in grayscale. On partial lithiation, the bulk Sb electrode expands and is imaged as shown in Fig.~\ref{Exp_F2}(b). (Note that the subfigure (b) in Fig.~\ref{Exp_F2} is relatively diffuse due to the low atomic weight of lithium). On complete lithiation to Li$_3$Sb, see Fig.~\ref{Exp_F2}(c), the bulk Sb electrode particle is observed to fracture into multiple fragments, with the most prominent crack extending from the center of the particle. Additional large cracks are visible along the edges, which suggests that the particle has undergone significant splitting during the electrochemical cycle. In Fig.~\ref{Exp_F2}(g) (black line), a typical bulk Sb particle expands by $\Delta \text{A} \approx 66\%$ (or equivalently an isotropic volume expansion of $\Delta \text{V}\approx 114\%$), which aligns with previously reported values for Sb cycled with lithium in the literature \cite{darwiche2012better}. 

\vspace{2mm}
\noindent Fig.~\ref{Exp_F2}(d-f) shows the analogous microstructural pathways for the nanoporous architectures of antimony electrodes during electrochemical cycling. The 2D TXM images show a representative NP-Sb particle in its pristine state that expands uniformly during lithiation with no significant cracking or fracture observed, when compared to the abrupt fragmentation observed in the bulk Sb electrode counterpart.

\vspace{2mm}
\noindent On complete lithiation, the NP-Sb particles expand by $\Delta \text{A}\approx$ 48$\%$ areally (averaged over six NP-Sb particles), which is 18 percentage points lower than the areal change in the bulk Sb electrode. This reduced expansion is associated with volume accommodation in the particles' porosity shown in Fig.~\ref{Exp_F2}, and demonstrates how the nanoporous structure effectively accommodates large structural changes accompanying the alloying reaction. The largest areal/volume expansion of the NP-Sb particles occurs during the initial stage of lithiation, which coincides with the Sb-to-Li$_2$Sb phase transformation observed by operando XRD (see Fig.~\ref{Exp_F1}(c), blue line). Below 0.75~V, the anode undergoes a Li$_2$Sb-to-Li$_3$Sb transformation (see Fig.~\ref{Exp_F1}(c), orange line), and the surface area (and net volume) of the NP-Sb particles changes only modestly when compared to the bulk particles, see Fig.~\ref{Exp_F2}(g). This behavior suggests that the internal porosity of NP-Sb accommodates part of the continued volume increase associated with alloying, thereby limiting changes in the particle’s external dimensions. We also note that the nanoporous architecture provides dual benefits: that is, it not only accommodates volume changes but also mitigates crack propagation, which collectively contributes to its improved structural reversibility during electrochemical cycling. We next investigate this interplay between phase transformation microstructures and electrode geometries (i.e., pores and cracks) using our continuum model. 

\section{Theory}
\label{sec:Theory}
\noindent Alloying electrodes offer high theoretical capacities, in which, guest species such as Li (or Na) react directly with another element (e.g., Sb). These reactions typically occur as a multi-step process and accommodate large concentrations of Li (or Na). For example, the alloying reaction of Sb electrodes in Li-ion batteries is typically a two-step process:
\begin{eqnarray}
    \mathrm{Sb + 2 Li^{+} + 2e^- \to Li_2Sb \qquad (Step~I)}\nonumber \\
    \mathrm{Li_2Sb + Li^+ + e^- \to Li_3Sb \qquad (Step~II)}
\end{eqnarray}

\noindent These addition reactions are characterized by abrupt changes in lattice symmetry and large-volume expansion. For example, inserting Li into Sb forms Li$_2$Sb with hexagonal symmetry (with large volume changes $\approx 90\%$) and continued lithiation forms Li$_3$Sb with cubic symmetry (with an additional $24\%$ volume change)\cite{baggetto2013intrinsic}. In this section, we formulate a continuum framework for an alloying electrode undergoing finite deformations. For step I of the alloying reaction (Sb$\to$Li$_2$Sb), we study the accompanying elastic-plastic deformation of the electrode analytically and derive the stresses at a phase boundary. For step II of the alloying reaction (Li$_2$Sb$\to$Li$_3$Sb), we construct a multi-physics variational model to predict the interplay between interfacial stresses, defects, and the evolving two-phase microstructure. 

\subsection{Analytical Model for an Expanding Spherical Electrode}
\noindent The Sb$\to$Li$_2$Sb alloying reaction is accompanied by an enormous $\approx~90\%$ volume change. Following the diffraction patterns of a partially lithiated Sb in Fig.~\ref{Exp_F1}(c), we note the coexistence of the Sb and Li$_2$Sb phases in crystalline forms despite the significant 90\% volume change accompanying the alloying reaction (see also Ref.~\cite{baggetto2013intrinsic}). We hypothesize that this large volume change is important in explaining the cracking of bulk Sb particles during phase change, and we analytically derive the interfacial stresses accompanying this reaction (see \ref{B} for details).

\begin{figure}[!ht]
\centering
\includegraphics[width=1.0\linewidth]{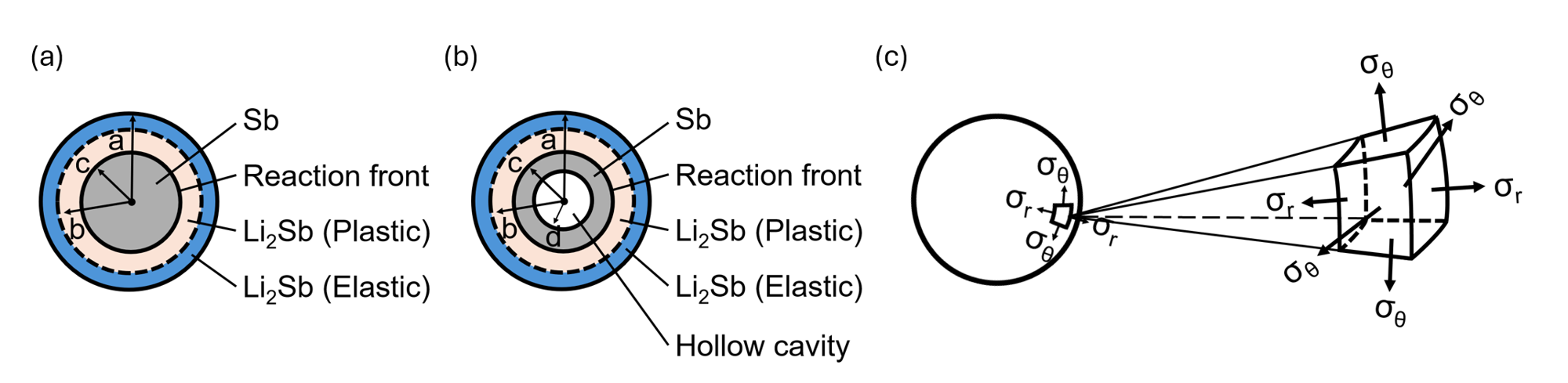}
\caption{Analytical study of lithiation in antimony particles. Schematic illustration of core-shell models for (a) bulk and (b) nanoporous antimony particles during the lithiation process. (c) Illustration of radial stress $\sigma_r$ and hoop stress $\sigma_{\theta}$ in spherical coordinates: note that crack propagation depends on the hoop stress at a given point.}
\label{Fig_Ana_model}
\end{figure}

\vspace{2mm}
\noindent Fig.~\ref{Fig_Ana_model} illustrates partially lithiated bulk and nanoporous spherical electrodes. These spherical particles with a rigid core and a hollow core are simplified representations of the experimentally synthesized bulk Sb and nanoporous Sb electrodes, respectively, see Section~\ref{sec:experiments}. The outer surfaces of these particles are unconstrained and expand freely during lithiation. In this outer shell, the region near the reaction front is assumed to be plastic (due to significant lattice misfit), whereas the region closer to the particle's free surface is assumed to be elastic (due to smaller lattice distortions). The inner core of the bulk Sb electrode is rigid and does not deform. Conversely, the core of the nanoporous Sb electrode is hollow and accommodates volume changes by shrinking into the pore. By assuming spherical geometries for the electrodes with isotropic volume expansion, we reduce the analytical problem to a 1D model in spherical polar coordinates ($r,\theta,\phi$). Fig.~\ref{Fig_Ana_model} (a-b) shows cross-sections of these partially lithiated electrodes of radius $r$ = a, an elastic-plastic interface at $r$ = b, and a phase boundary positioned at $r$ = c. In the nanoporous electrode, the pore is a sphere of radius $r$ = d. The balance of forces acting on a material element at any position $r$ requires that \cite{hill1998mathematical}:
\begin{align}
    \frac{\mathrm{d} \sigma_r(r)}{\mathrm{d} r} +  2\frac{\sigma_r(r)-\sigma_\theta(r)}{r} = 0.
    \label{Force_equilibrium}
\end{align}

\vspace{2mm}
\noindent Eq. (2) represents the mechanical equilibrium condition for an expanding spherical electrode, in which $\sigma_r$ represents the radial stress and $\sigma_\theta$ the hoop stress acting on a representative material element (See Fig.~\ref{Fig_Ana_model} (c)). These stresses have explicit analytical forms in the spherical polar coordinates, and we solve these stresses for the two boundary conditions corresponding to the bulk and nanoporous electrode geometries, respectively. In the elastic region, the radial and hoop stress components in spherical coordinates are given by the Lam\'{e} solution \cite{hill1998mathematical}:
\begin{align}
\sigma_r(r) & = A - \frac{B}{r^3} \label{Eq: Radial Stress}\\
\sigma_{\theta}(r) &= A + \frac{B}{2r^3}, \label{Eq: Hoop Stress}
\end{align}

\vspace{2mm}
\noindent In the plastic region, the stress state is constrained by the yield condition, $\sigma_\theta(r) -\sigma _r(r)=\sigma _Y$. Substituting this condition into Eq.~\ref{Force_equilibrium} gives the radial and hoop stresses in the following forms:
\begin{equation}
    \begin{aligned}
        & \mathrm{\sigma_{r}}(r) = 2\sigma_Y \ln{r} + C\\
        & \mathrm{\sigma_{\theta}}(r) = \sigma_Y + 2\sigma_Y \ln{r} + C
    \end{aligned}
    \label{Eq: Plastic stress}
\end{equation}

\vspace{2mm}
\noindent Here, $A$, $B$, and $C$ are integration constants determined by the corresponding boundary and continuity conditions. A detailed description of these conditions is presented in \ref{B}.

\begin{figure}[!ht]
\centering
\includegraphics[width=1.0\linewidth]{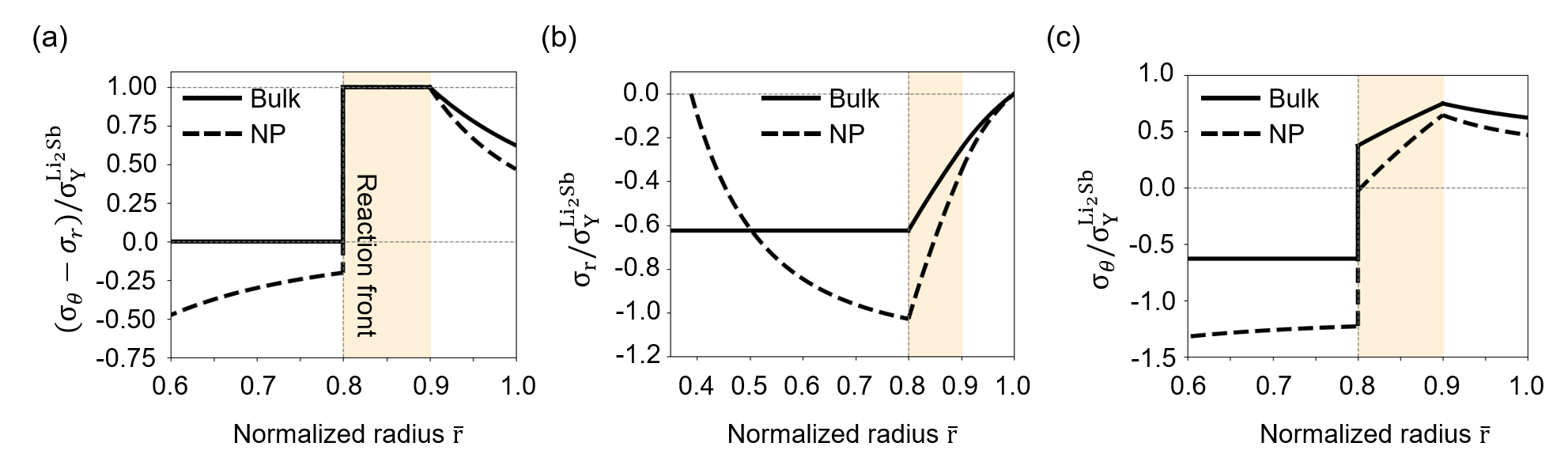}
\caption{Analytical solutions of stresses in lithiated antimony particles: (a) the biaxial stress, $\sigma_\theta-\sigma_r$, (b) the radial stress, $\sigma_r$, and (c) the hoop stress, $\sigma_\theta$, in bulk and nanoporous particles when reaction front is at $\bar{r}=0.8$. In (c), the nanoporous particle exhibits consistently lower hoop stress than the bulk particle, demonstrating the role of the pore in accommodating reaction-induced deformation. The radial coordinate is normalized such that the outer particle surface corresponds to $\bar{r}=1$, while the reaction front corresponds to $\bar{r}=0.8$ for both particle geometries. The shaded-orange regions indicate the plastic zone.}

\label{fig:lithiation}
\end{figure}

\noindent Figs.~\ref{fig:lithiation}(a)--(c) show the biaxial stress, $\sigma_\theta-\sigma_r$, the radial stress, $\sigma_r$, and the hoop stress, $\sigma_\theta$ in the bulk and nanoporous electrode particles. These plots together show that the tensile hoop stress concentrated at the reaction front is the leading cause of failure in bulk Sb electrodes. These stresses are reduced in nanoporous electrode geometries and contribute to their improved cycling. Fig.~\ref{fig:lithiation}(a) shows the onset of plastic deformation at $\bar{r}=0.9$ at which the biaxial stress is maximum and equal to the yielding criteria, $\sigma_\theta-\sigma_r=\sigma_\mathrm{Y}^{\mathrm{Li_2Sb}}$. This plastic zone is in the vicinity of the reaction front and is highlighted by the shaded-orange regions in Figs.~\ref{fig:lithiation}(a-c). Fig.~\ref{fig:lithiation}(b) shows that the core of bulk Sb particle is under uniform hydrostatic compression during cycling, which is relieved in nanoporous electrode geometries at the traction-free pore surfaces ($\bar{r}$ $\approx$ 0.4). For the hoop stress shown in Fig.~\ref{fig:lithiation}(c), both electrode geometries exhibit tensile stresses at the outer surface, with the NP electrode showing a lower hoop stress than the bulk electrode. This hoop stress decreases upon entering the plastic zone, and is  compressive and likely suppresses crack propagation in the unreacted region. A similar trend has been reported \cite{zhao2011concurrent}.

\vspace{2mm}
\noindent We next analyze the effect of interfacial hoop stresses on a pre-existing crack in Sb electrodes. According to Griffith's theory of fracture, pre-existing cracks propagate when the elastic energy stored in the material bulk exceeds the energy required to create new surfaces. This balance between the energy terms gives a fracture criterion:
\begin{equation}
    \begin{aligned}
        &  a=\frac{G_cE}{\pi \sigma_\theta^2}.
    \end{aligned}
    \label{Eq: Fracture Criterion eq}
\end{equation}

\noindent In Eq.~(\ref{Eq: Fracture Criterion eq}), the critical size of a pre-existing crack to propagate is a function of the material's critical energy release rate $\mathcal{G}_c$, Young's modulus $E$, and we focus on fracture caused by the tensile hoop stress $\sigma_\theta$ acting at the crack. With $E = 56$ GPa~\cite{jain2013commentary}, $\mathcal{G}c = 12~\mathrm{J/m^2}$ \cite{zhao2011inelastic}, assuming $\sigma_\mathrm{Y}=$ 1 GPa \cite{zhao2012reactive} for Li$_2$Sb (comparable to fully lithiated silicon alloying anode), we estimate that pre-existing cracks with $a\ge 0.37~\mu\mathrm{m}$ will propagate in bulk electrodes under maximum hoop stress of $\sigma_\theta =0.76\sigma_\mathrm{Y}$ when reaction front is at $\bar{r}=0.8$. Our analysis for the nanoporous electrode predicts that relatively larger cracks of about $a\ge 0.56~\mu\mathrm{m}$ will propagate under the maximum hoop stress of approximately $\sigma_\theta\approx 0.62\sigma_\mathrm{Y}$. This increase in the critical crack length supports the observation of fewer cracks reported in the nanoporous systems.

\subsection{A Variational Model for Finite Volume Changes}
\noindent The second step of the alloying reaction, Li$_2$Sb $\to$ Li$_3$Sb is accompanied by finite volume changes (up to 24\%), and is commonly characterized in experiments to have reduced cracking. This structural failure is further reduced in nanoarchitected electrodes, with the pores accommodating the finite volume changes and relieving internal stresses. Here, we outline a theoretical framework for step II of the phase transformation process, in which a primary electrode particle is in contact with an electrolyte reservoir. We apply this model in the remainder of this work, to investigate the interplay between defects and a moving phase boundary in bulk and nanoporous electrodes. The model is general and can be adapted to any alloying anode (with moderate volume changes), but for our purposes, we use Li$_2$Sb $\to$ Li$_3$Sb as a representative example.

\vspace{2mm}
\noindent We begin with a generalized diffusion-reaction phase-field model derived in Ref.~\cite{afshar2021thermodynamically} and build on this framework to predict the nonlinear stress-strain behavior in alloying electrodes. We use the neo-Hookean material law to predict the large deformations of the electrodes accompanying an alloying reaction. We calibrate our modeling framework for the Li-Sb system and investigate the stress accumulation and volume accommodation in the electrodes during phase transformation.

\begin{figure}[ht]
    \centering
    \includegraphics[width=0.7\textwidth]{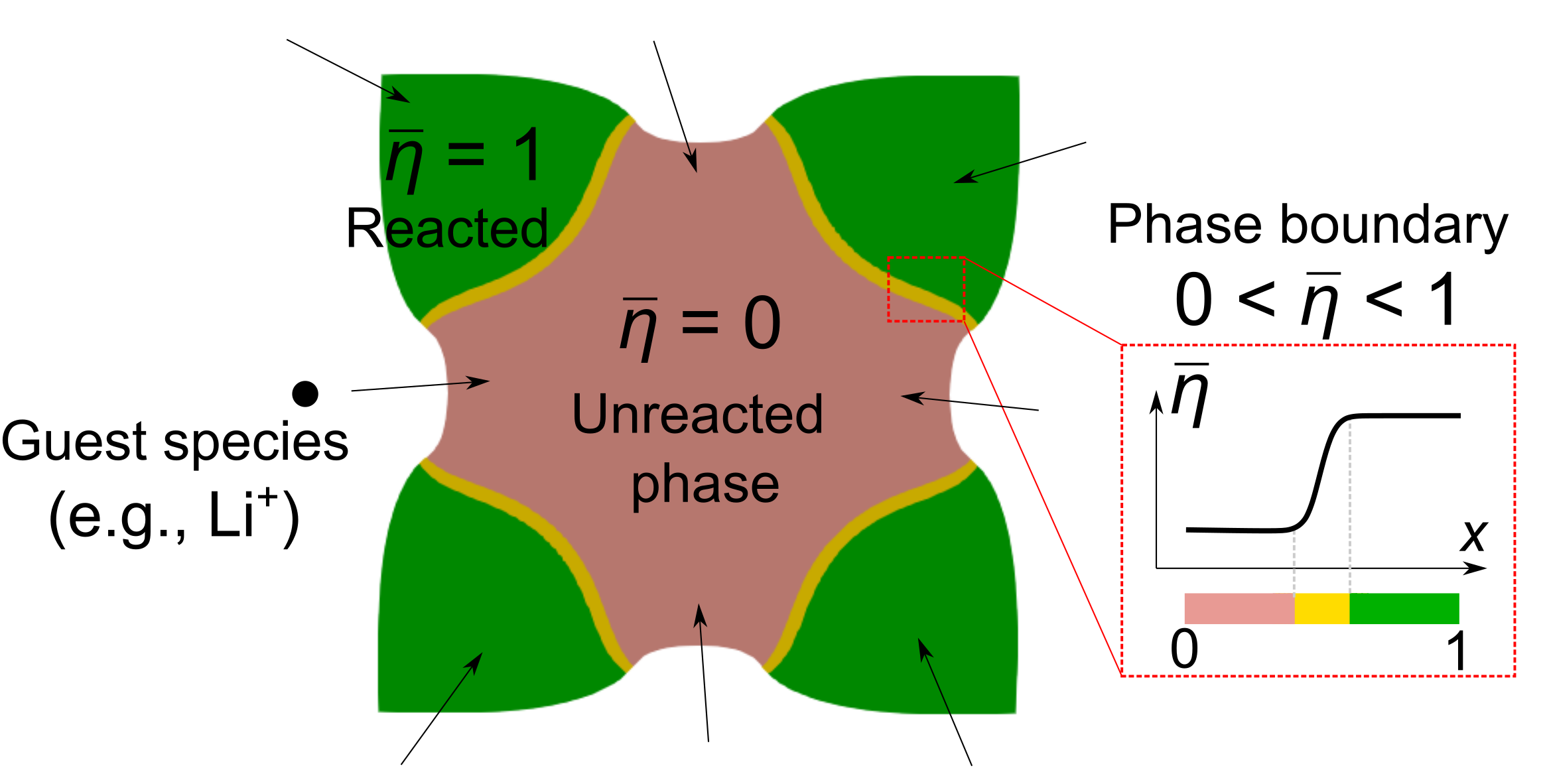}
    \caption{Schematic illustration of an electrode particle undergoing an alloying reaction. The guest species (e.g., Li$^{+}$) reacts with a pristine electrode to form a reacted phase. We denote the extent of this reaction by an order parameter $\bar{\eta}$ and a phase interface $0<\bar{\eta}<1$ forms phenomenologically between the two phases. Both the reacted and pristine phases serve as hosts for the diffusing Li-species $c$. Note that the alloying reaction is accompanied by significant volume changes.}
    \label{fig:diffusion-reaction schematic}
\end{figure}

\vspace{2mm}
\noindent In our model, the order parameter $0 \le \bar{\eta} \le 1$ represents the extent of the alloying reaction (with $\bar{\eta} = 1$ representing the fully-reacted phase), as shown schematically in Fig.~\ref{fig:diffusion-reaction schematic}, and $0 \le \bar{c} \le 1$ represents the normalized Li-concentration diffusing in the reacted/pristine phases. We describe the alloying reaction using linearized kinetics:
\begin{align}
    \frac{\partial \eta}{\partial t} &= -\frac{\mathrm{R_0}\eta_{\mathrm{max}}}{\mathrm{RT}}[\tilde \mu-\mu]
    \label{eq:reaction kinetics}
\end{align}

\noindent with $\mathrm{R}_0$ as the reaction constant, and $\tilde{\mu}$ and $\mu$ represent chemical potentials. The values of all material constants, including the Gas constant $\mathrm{R}$ and room temperature $\mathrm{T}_0$, are listed in Table~\ref{tab:material constants} of the ~\ref{A}. Eq.~(\ref{eq:reaction kinetics}) describes a chemical reaction process in which a driving force---defined as the difference in chemical potentials between participating species---drives the alloying reaction. Lithium diffusion in the reacted and pristine phases is described using:
\begin{align}
    \frac{\partial c}{\partial t} = \nabla\cdot[\mathbf{M}(c)\nabla\mu] - \frac{\partial \eta}{\partial t}
    \label{eq:diffusion/reaction}
\end{align}

\noindent Here, $\mathbf{M}(c) = \frac{\mathrm{D_0}}{\mathrm{RT}}c(1-\bar{c})\mathbf{I}$ is the mobility tensor (with $\mathrm{D_0}$ as the diffusion constant) and a gradient of the chemical potential $\nabla \mu$ acts as a driving force for the diffusion process. Eq.~(\ref{eq:diffusion/reaction}) describes a coupled diffusion-reaction kinetics in which both the mass flux and the reaction rate govern the diffusion process during the alloying reaction.

\vspace{2mm}
\noindent We describe the chemical potentials $\mu, \tilde{\mu}$ in a variational sense. That is, we define $\mu$ as the variational derivative of the total free energy of the system $\Psi$ ($\Psi=\int_\mathcal{B}\psi~\mathrm{d}\mathbf{x}$) with respect to $c$:
\begin{align}
    \mu &= \frac{\delta \Psi}{\delta c},
    \label{eq:diffusion chemical potential}
\end{align}

\vspace{2mm}
\noindent and the reaction chemical potential $\tilde{\mu}$ as the partial derivative of the free energy density with respect to $\eta$ and is dependent on the Mandel stress $\bar{\mathbf{M}}$ (that we describe later) and the direction of reaction-induced deformation $\mathbf{N}$:
\begin{align}
    \tilde{\mu} &= \frac{\partial \psi}{\partial \eta} - \mathrm{J}_0 \bar{\eta} \bar{\mathbf{M}}:\mathbf{N}-\nabla \cdot \frac{\kappa}{\eta_{\mathrm{max}}}\nabla \bar{\eta}.
    \label{eq:reaction chemical potential}
\end{align}

\vspace{2mm}
\noindent In our modeling framework, we assume the total deformation of the host electrode to be from the alloying reaction and neglect the diffusion-induced deformation of the host electrode. We define $\mathbf{N} = \frac{\Omega}{3(1+\Omega\eta)}\mathbf{I}$ as an isotropic tensor and as a function of the partial molar volume $\Omega$ corresponding to the reaction-induced deformation of the electrode. 

\vspace{2mm}
\noindent The total free energy of the system $\Psi$ accounts for the thermodynamic contributions from the extent of alloying reaction $\psi_{\mathrm{rxn}}$, the interfacial free energy that penalizes interfaces $\psi_{\mathrm{grad}}$, the chemical free energy of mixing of the diffusion species $\psi_{\mathrm{chem}}$, and accounts for the elastic energy arising from the reaction-induced finite deformation of the host electrode $\psi_{\mathrm{elas}}$.  The total free energy of an electrode particle $\mathcal{B}$ in three-dimensional Euclidean space $\mathbb{R}^3$ is given by:
\begin{align}
    \Psi &=\int_\mathcal{B} \Bigl\{ \underbrace{E_a \bar{\eta}^2(1-\bar{\eta})^2 + \eta \mu_{\eta} }_\text{$\psi_{\mathrm{rxn}}$}+\underbrace{\frac{1}{2}\kappa|\nabla\bar{\eta}|^2}_\text{$\psi_{\mathrm{grad}}$}\nonumber\\
    & + \bar{\eta} \mathrm{c_{max}}\{\mu_{\eta}\bar{c} + \mathrm{RT} [\bar{c} \mathrm{ln} \bar{c} + (1-\bar{c}) \mathrm{ln}(1-\bar{c})]\} \nonumber\\
    & + \underbrace{(1-\bar{\eta})\mathrm{c_{max}}\{\mu_{\alpha}\bar{c} + \mathrm{RT} [\bar{c} \mathrm{ln} \bar{c} + (1-\bar{c}) \mathrm{ln}(1-\bar{c})]\}}_\text{$\psi_{\mathrm{chem}}$} \nonumber\\
    & + \underbrace{{\mathrm{J}_0[\frac{1}{2}\mathrm{G}(\mathbb{I}_1-3) + \frac{1}{2}\mathrm{\lambda}(\mathrm{ln}\mathbb{I}_3)^2 - \mathrm{G} \mathrm{ln}\mathbb{I}_3]}}_\text{$\psi_{\mathrm{elas}}$}\Bigl\} \mathrm{d}\mathbf{x}
    \label{eq:free energy}
\end{align}
\noindent with invariants $\mathbb{I}_1 = \mathrm{tr}(\bar{\mathbf{F}}^{\top}\bar{\mathbf{F}})$ and $\mathbb{I}_3 = \sqrt{\mathrm{det}(\bar{\mathbf{F}}^{\top}\bar{\mathbf{F}})}$ \cite{holzapfel2002nonlinear,bucci2016formulation}. The deformation gradient $\bar{\mathbf{F}}$ corresponds to electrode distortions away from the stress-free states and is described in detail below.

\begin{figure}[ht]
    \centering
    \includegraphics[width=\textwidth]{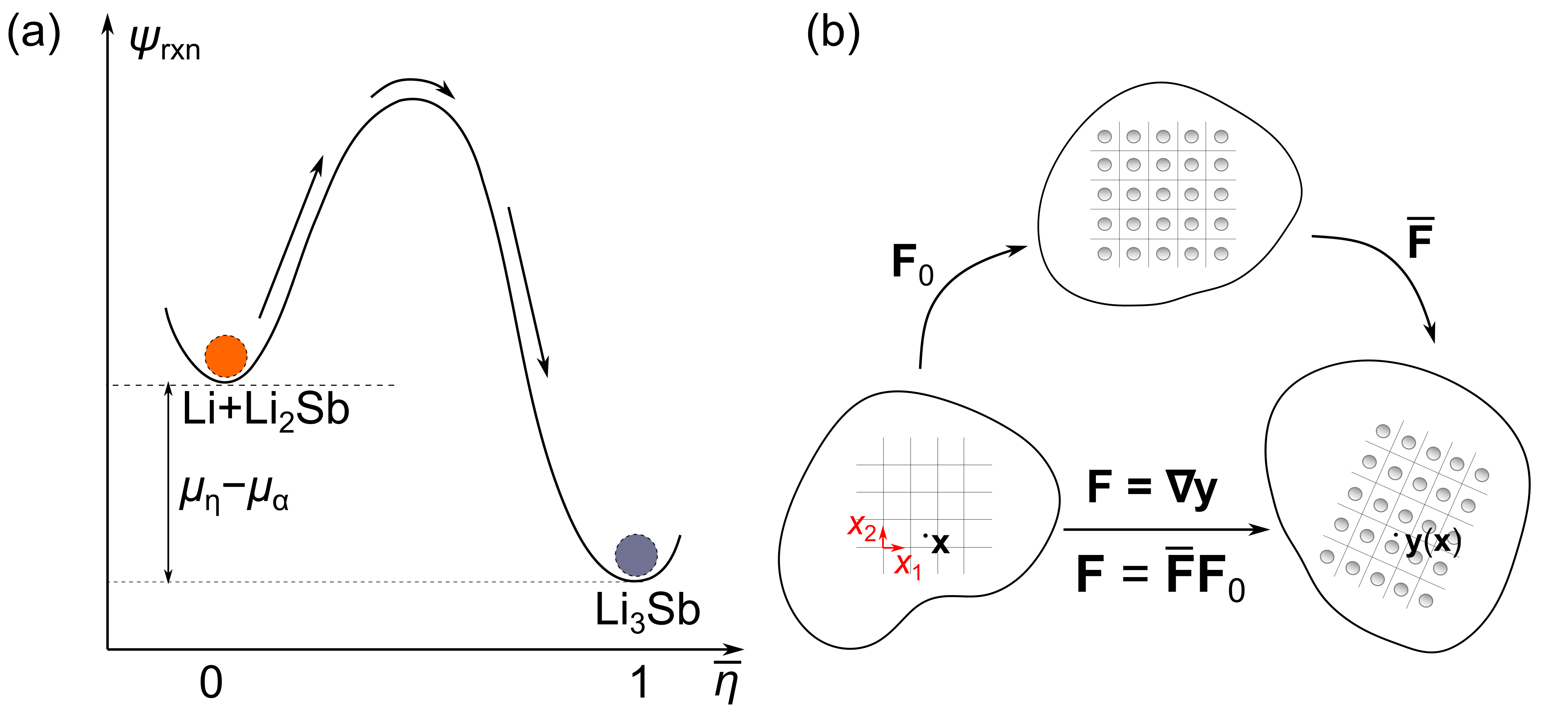}
    \caption{(a) We construct a free energy landscape as a function of the order parameter $\bar{\eta}$ with the wells corresponding to the Li$_2$Sb phase $\bar{\eta} = 0$ and the Li$_3$Sb phase $\bar{\eta}=1$, respectively. We calibrate the energy barrier height with the activation energy $E_a$ for Li$^+$ hopping in Li$_2$Sb, and the position of the two wells represents the chemical potential difference driving the phase change $\mu_{\eta} - \mu_{\alpha}$. Here, $\mu_\alpha$ is the chemical potential of the unreacted phase, and $\mu_{\eta}$ is the chemical potential of the reacted phase. (b) A schematic illustration of the kinematic decomposition of the deformation gradient used in our model: During phase change, a material point $\mathbf{x}$ in the reference configuration deforms affinely as a function of $\eta$ to a stress-free intermediate configuration. We denote this change in configuration by $\mathbf{F}_0(\eta)$ and attribute the elastic energy stored in the system as deviations from this stress-free state. These deviations are represented by $\bar{\mathbf{F}}$ and the deformation gradients are related as $\mathbf{F} = \nabla \mathbf{y} = \bar{\mathbf{F}}\mathbf{F}_0$.}
    \label{fig:reaction}
\end{figure}

\vspace{2mm}
\noindent Potential energy of chemical reaction  $\psi_{\mathrm{rxn}}$ in Eq.~(\ref{eq:free energy}) describes a double-well energy landscape as a function of the normalized reaction order parameter $\bar{\eta}$, see Fig.~\ref{fig:reaction}(a). The chemical activation energy $E_a$ governs the height of the reaction energy barrier. The second term sets the reference chemical potential to $\mu_{\eta}$, representing the reference potential of the species (i.e., Li$^+$) when it is chemically reacted and part of the new compound (i.e., Li$_3$Sb).

\vspace{2mm}
\noindent A regular solution model is used to describe the chemical free energy of mixing of the diffusing species $0 \leq \bar{c} \leq 1$ with $\mu_\alpha$ and $\mu_\eta$, respectively, representing the reference potentials of the diffusion species (i.e., Li$^+$) in the unreacted (Li$_2$Sb) and reacted (Li$_3$Sb) phases of the host electrode.

\vspace{2mm}
\noindent The elastic energy stored in the system is described using the neo-Hookean material law and penalizes deformations that deviate from energy-minimizing states. For example, in Fig.~\ref{fig:reaction}(b) $\mathbf{x}$ is the position of a point on the reference configuration $\mathcal{B}$ of the electrode. We describe the deformation of this electrode as a function $\mathbf{y}: \mathcal{B} \to \mathbb{R}^3$ 
in which $\mathbf{y}(\mathbf{x})$ denotes the position of the point $\mathbf{x}$ in the deformed configuration. We define a deformation gradient with respect to the reference configuration as $\mathbf{F} = \nabla\mathbf{y}$.

\vspace{2mm}
\noindent During the alloying reaction, the electrode deforms in an affine manner as a function of $\eta$. These intermediate deformations are stress-free and the corresponding deformation gradients $\mathbf{F}_0(\eta)$ represent a change in the reference configuration of the electrode, see Fig.~\ref{fig:reaction}(b). Any deformation away from the stress-free states, represented by the deformation gradient $\bar{\mathbf{F}}$, contributes to a finite elastic energy stored in the system. These deformation gradients are related to the net deformation of the electrode via a multiplicative decomposition $\mathbf{F} = \bar{\mathbf{F}}\mathbf{F}_0$, see Fig.~\ref{fig:reaction}(b) \cite{ogden1997non}. The elastic energy is frame invariant and is defined as a function of a right Cauchy-Green tensor $\bar{\mathbf{C}}=\bar{\mathbf{F}}^{\top}\bar{\mathbf{F}}$. We assume a neo-Hookean material law and describe the non-linear stress-strain behavior of the alloying electrode undergoing large deformations as shown in Eq.~(\ref{eq:free energy}). The coefficients $\mathrm{G}$ and $\lambda$ correspond to the Lam\'e parameters and are related to the shear and bulk modulus of the host electrode, respectively. The specific values of the material constants are listed in Table \ref{tab:material constants} and details on the material kinematic assumptions are discussed in \ref{C}. We implement the model in the open-source MOOSE finite-element, multiphysics framework, and adopt the mixed formulation in which $\mu$, $\eta$ as well as $c$ are regarded as primary unknowns. Further details on the material constants and numerical implementations used in our calculations are described in ~\ref{A} and ~\ref{C}, respectively.

\vspace{2mm}
\noindent In our calculations we use four types of stress tensors that relate forces with areas in reference, intermediate, or deformed configurations. All stress tensors have the standard definition in continuum mechanics, and for our purposes, we list the relations between different stress tensors and describe the mechanical equilibrium conditions solved in our model:

    \begin{itemize}
        \item The first Piola-Kirchhoff stress tensor $\mathbf{P}$ is derived from the local dissipation inequality ($\mathbf{P} = \frac{\partial \psi}{\partial \mathbf{F}}$) and satisfies the mechanical equilibrium condition:
        \begin{eqnarray}
        \nabla\cdot\mathbf{P}^{\top} = 0.\label{eq:force balance}
        \end{eqnarray}
        \item The first Piola-Kirchhoff stress tensor is related to the symmetric Cauchy stress tensor $\boldsymbol{\sigma}$ and the Jacobian determinant $\mathrm{J = det}\mathbf{F}$ as:
        \begin{eqnarray}
        \mathbf{P} = \mathrm{J}\boldsymbol{\sigma}\mathbf{F}^{-\top}.
        \end{eqnarray}
        The first Piola-Kirchhoff stress is not necessarily a symmetric tensor.
        \item For our numerical calculations, we define a symmetric stress tensor (referred to as the second Piola-Kirchhoff stress $\bar{\mathbf{S}}$) on the stress-free intermediate configuration of the electrode. This symmetric stress accounts for deformation gradients $\bar{\mathbf{F}}$ away  from the stress-free configuration and is related to the Cauchy stress tensor $\mathbf{\sigma}$ as:    
        \begin{align}
        \bar{\mathbf{S}} = \bar{\mathrm{J}} \bar{\mathbf{F}}^{-1}\boldsymbol{\sigma}\bar{\mathbf{F}}^{-\top}.
        \label{eq:PK2 stress}
        \end{align}
        In Eq.~(\ref{eq:PK2 stress}) $\bar{\mathrm{J}}=\mathrm{det}\bar{\mathbf{F}}$ is the Jacobian determinant with respect to $\bar{\mathbf{F}}$.
        \item Finally, the Mandel stress tensor is defined on the stress-free intermediate configuration as:
        \begin{align}
        \bar{\mathbf{M}} = \bar{\mathbf{C}}\bar{\mathbf{S}},
        \end{align}
        in which $\bar{\mathbf{C}}$ is the right Cauchy-Green tensor described above.
    \end{itemize}

\section{Results} \label{sec:results}
\noindent Using the Li$_2$Sb $\to$ Li$_3$Sb alloying reaction as a representative example, we investigate the interplay between phase transformation pathways and defects, such as pores or cracks, starting from a Li$_2$Sb electrode. Our results show that the microstructural pathways during phase change affect internal stresses and their accumulation at crack tips. At critical values, these internal stresses propagate cracks leading to material failure. The large volume changes ($\sim 24\%$) of the alloying electrode can be accommodated in pores, however, phase boundaries, sharp corners, and slender walls generate significant internal stresses. Finally, we demonstrate a potential application of our model as an electrode design tool in which \textcolor{black}{geometric instabilities, such as bending}, are harnessed to effectively accommodate large deformations with small internal stresses.

\vspace{2mm}
\noindent In our computations, we assume plane-strain conditions and model a 2D domain representing a primary particle of size $5\ \mu\mathrm{m} \times 5\ \mu\mathrm{m}$, subject to a constant flux on the surface. We introduce defects by selectively trimming elements and apply the Li flux to the surfaces of defects with $>3\%$ surface area. The latter is motivated by experimental observations that electrolyte seeps through larger pores and reduces the diffusion length \cite{mcdowell201325th}. For further details on the numerical implementation, please see the ~\ref{C}.% and our code on the \textcolor{gray}{Open Science Framework \cite{}.}

\begin{figure}[ht!]
    \centering
    \includegraphics[width=\textwidth]{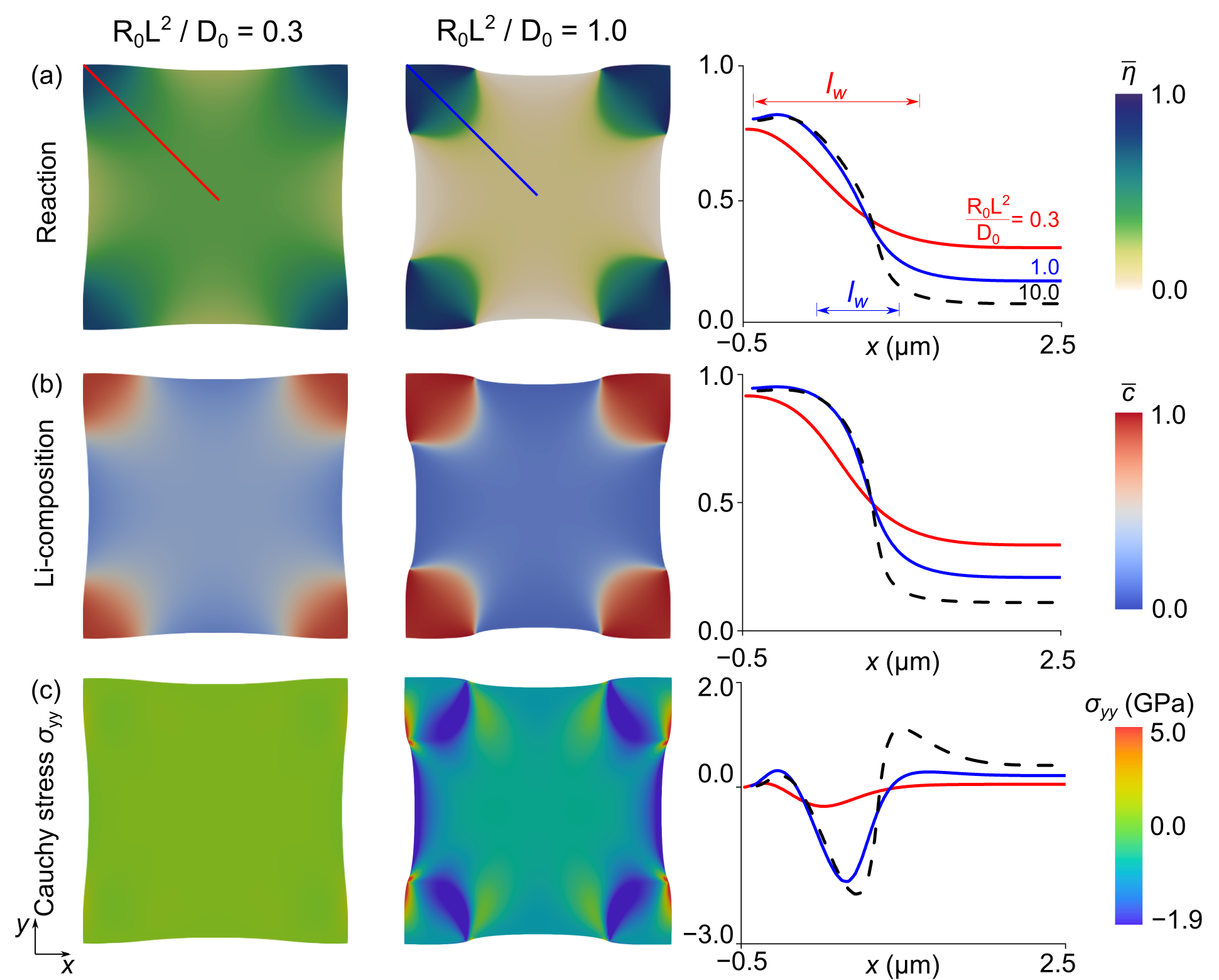}
    \caption{\textcolor{black}{Effect of diffusion and reaction kinetics, characterized by $\mathrm{\frac{R_0L^2}{D_0}}$, on the phase-boundary morphology and internal stresses in the Li$_2$Sb/Li$_3$Sb alloying reaction. (a) The reaction degree, (b) Li concentration, and (c) Cauchy stress component $\sigma_{yy}$ are shown. Each subfigure also includes plots showing the distribution of these variables across a phase boundary. For $\mathrm{\frac{R_0 L^2}{D_0}} =0.3$, the phase boundary is diffuse, however, with increasing reaction rates the phase boundary becomes sharp, see $\mathrm{\frac{R_0 L^2}{D_0}} \ge 1.0$. As the reaction rate further increases $\mathrm{R_0 \gg D_0}$, the width of the interface remains constant; however, significant stresses accumulate at the phase boundary. This balance between diffusion kinetics and reaction rates generates a range of tensile stresses at the interface, as shown in subfigure (c), which contributes to the propagation of pre-existing cracks in the electrode.}}
    \label{Fig2}
\end{figure}

\subsection{Internal stresses during alloying reaction}

%\vspace{2mm}
\noindent Fig.~\ref{Fig2} shows the effect of coupled diffusion-reaction kinetics on the phase-boundary morphology and internal stresses in Li$_2$Sb. We introduce a non-dimensional quantity $\mathrm{\frac{R_0 L^2}{D_0}}$ to analyze the effect of competing reaction and diffusion rates in a two-phase microstructure (Li$_2$Sb/Li$_3$Sb) at $\sim 33\%$ state of charge (SOC). When Li-diffusion in individual phases is faster than the Li$_2$Sb to Li$_3$Sb reaction ($\mathrm{\frac{R_0 L^2}{D_0}} = 0.3$), the phase boundary is diffuse ($l_w = 1.5~\mu\mathrm{m}$) and the maximum stress at the interface is $\sigma_{yy} = 0.4$ GPa, see the first column in Fig.~\ref{Fig2}(a-c). However, with increased reaction rates, the lithiation enters the diffusion-limited regime (i.e., the reaction kinetics are much faster than the diffusion kinetics). As $\mathrm{\frac{R_0 L^2}{D_0}}$ approaches 1, the phase boundary evolves into a sharper interface ($l_w = 0.7\mu\mathrm{m}$) generating significant interfacial stresses ($-1.9\ \mathrm{GPa} < \sigma_{yy} <$ 5 GPa), see stress distribution in Fig.~\ref{Fig2}(c). At very large reaction rates $\mathrm{\frac{R_0 L^2}{D_0}}=10$, the phase boundary thickness approaches the sharp interface limit, and the large volume changes across this interface generate enormous interfacial stresses.

\begin{figure}[ht!]
    \centering
    \includegraphics[width=0.9\textwidth]{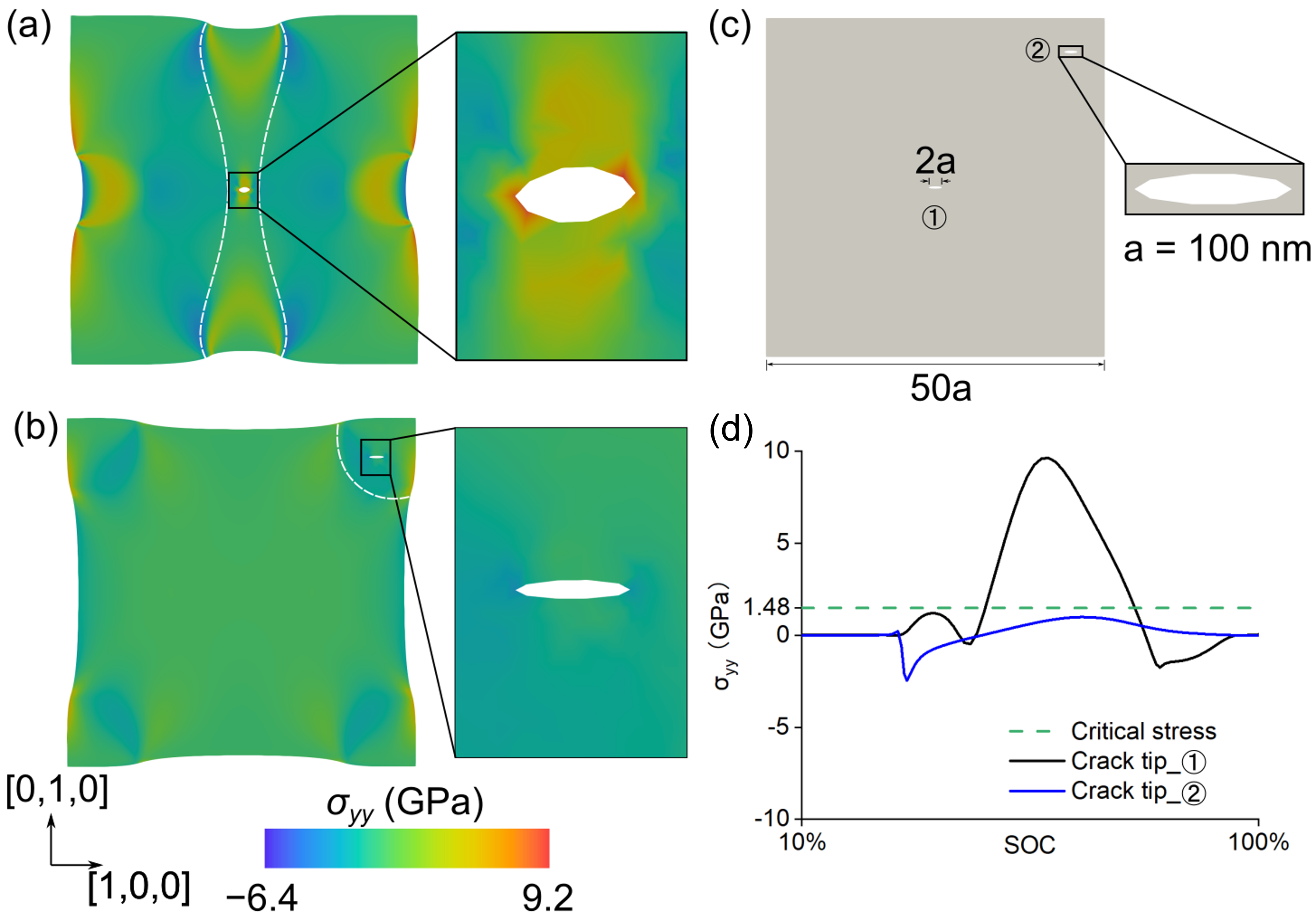}
    \caption{\textcolor{black}{We analyze the stresses in the vicinity of two representative lenticular-shaped cracks, positioned at $\textcircled{1}$ and $\textcircled{2}$, respectively, in a Li$_2$Sb particle. (a) At $\textcircled{1}$, the crack is positioned at the phase boundary separating the Li$_2$Sb/Li$_3$Sb regions. The enormous stresses at the phase boundary coupled with significant volume changes accompanying the alloying reaction generate tensile stresses of up to $\sigma_{yy} = 9.2\ \mathrm{GPa}$ at the crack tips. This peak stress exceeds the critical threshold of 1.48 GPa necessary for crack propagation in Sb. (b) By contrast, the crack $\textcircled{2}$ located within a fully transformed phase of Li$_3$Sb undergoes compression of up to $\sigma_{yy} = -3.87\ \mathrm{GPa}$. These compressive stresses close the crack opening and delay the onset of fracture. (c-d) We track the peak stress values at both $\textcircled{1}$ and $\textcircled{2}$ during lithiation and note that the position of cracks within the particle plays an important role in the structural reversibility of the electrode particle. The higher stresses at crack $\textcircled{1}$ make it more prone to failure during electrochemical cycling. Note, all calculations were done with a Li-flux of 0.5 C-rate.}}
    \label{fig: cracks}
\end{figure}

\vspace{2mm}
\noindent With repeated cycling, these phase boundaries interact with defects, such as pre-existing cracks, and generate internal stresses. Fig.~\ref{fig: cracks}(a-d) shows the stress distributions around lenticular-shaped cracks ($a = 0.1\ \mu\mathrm{m}$) positioned at two different locations within the electrode particle (marked as $\textcircled{1}$ and $\textcircled{2}$ in Fig.~\ref{fig: cracks}(c)). We use these pre-existing cracks as representative examples of defects that are located at the phase boundary and embedded within a fully transformed region of the electrode, respectively. We apply a Li flux (0.5 C-rate) on all surfaces with $\mathrm{\frac{R_0 L^2}{D_0}} = 1.0$. At SOC = 66.3$\%$ the abrupt volume changes across the phase boundary generate compressive stresses $\sigma_{yy} = -6.4$ GPa in the core of the unlithiated phase, and tensile stresses of $\sigma_{yy} = 9.2$ GPa at the crack tips, see Fig.~\ref{fig: cracks}(a). By contrast, the crack embedded within the lithiated phase Li$_3$Sb is under compression $\sigma_{yy} = -3.87$ GPa, see Fig.~\ref{fig: cracks}(b).

\vspace{2mm}
\noindent The tensile stresses at the crack tips $\sigma_{yy} = 9.2$ GPa exceed the critical stress value necessary to propagate cracks in antimony. By Griffith's criterion, a crack of size $a = 0.1\mu\mathrm{m}$ in an Sb particle of size $\sim5~\mu\mathrm{m}$ fails at a critical stress value of $\sigma_c = \sqrt{\frac{\mathcal{G}_c E}{\pi a}} \approx 1.48$ GPa. In the absence of detailed experimental measurements on mechanical properties of Sb, we have assumed a critical energy release rate $\mathcal{G}_c = 12~\mathrm{J/m^2}$ that is comparable to other anode materials such as single-crystal silicon electrodes \cite{shi2016failure}, and Young's modulus of $E = 56\ \mathrm{GPa}$ from Materials Project \cite{jain2013commentary} for Li$_3$Sb. The small critical stresses for Li$_3$Sb and the large volume expansion during phase change would explain the common observation of cracked electrodes during lithiation in our experiments in section~\ref{sec:experiments}. The cracks embedded in the fully transformed region are under compressive stresses that close the crack opening and delay the onset of fracture. These two representative computations highlight the importance of the relative position of cracks in a two-phase microstructure of the electrode particle. 

\subsection{Volume Accommodation in Pores}
\noindent In our experiments (section~\ref{sec:experiments}), we showed that nanostructured electrodes demonstrate improved cycling performance, which is widely attributed to the nanoporous geometry accommodating the large volume changes associated with the reversible alloying reaction \cite{an2018micron,li2021polyaniline}. We then analytically showed (section~\ref{sec:Theory}) that these electrodes contain nano-sized pores or interconnected pores into which the electrode swells, thereby reducing the net volume expansion and interfacial stresses. In this subsection, we identify the phase transformation pathways in nanoporous electrodes and establish the interplay between concentration flux (internal driving force) and porous geometries during alloying reactions. 

\begin{figure}[ht]
    \centering
    \includegraphics[width=\textwidth]{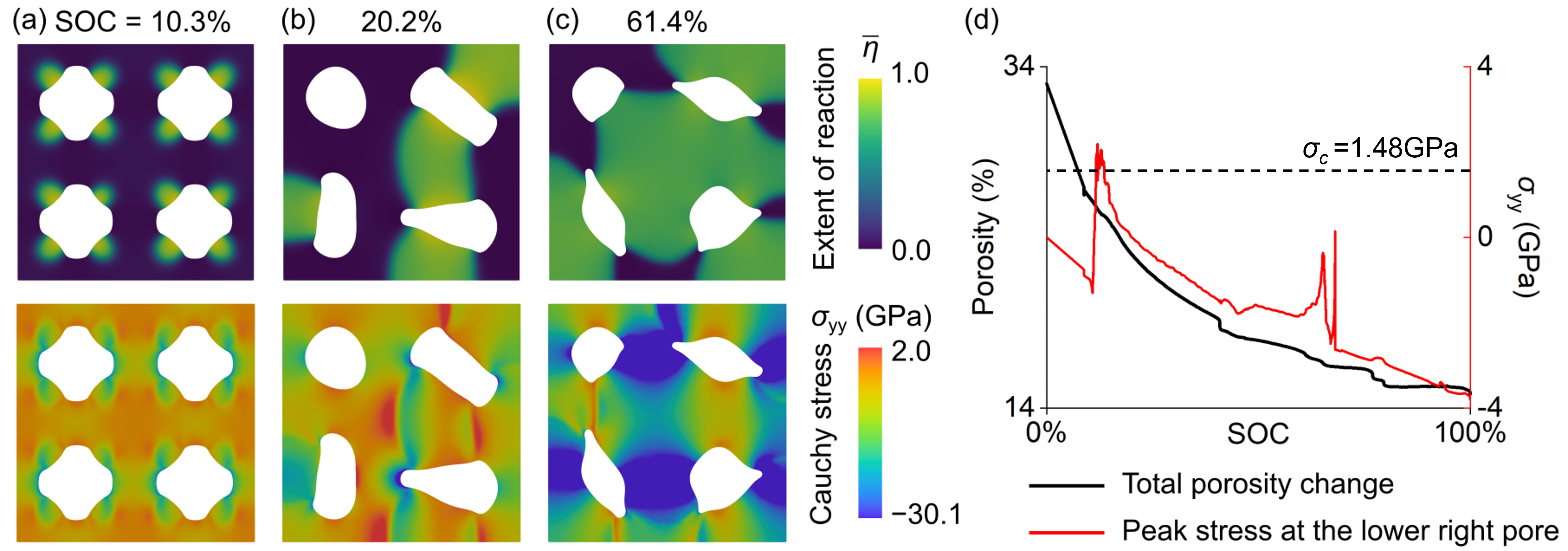}
    \caption{(a-c) A phase transformation pathway in a representative nanoporous Li$_2$Sb electrode. (a-c; Top row) At SOC = 10.3$\%$, a Li$_3$Sb phase nucleates around pores and with continued lithiation a two-phase microstructure evolves in the electrode. This Li$_2$Sb $\to$ Li$_3$Sb transformation involves significant deformation of the pores. At SOC $=61.4\%$, the pores accommodate a majority of the volume change of the electrode during the alloying reaction. (a-c; Bottom row) During lithiation, the stresses around the pore walls are primarily compressive with a peak value of $\sigma_{yy}$ = $-$10.1 GPa. The phase boundaries show tensile stress of up to $\sigma_{yy}$ = 2.0 GPa during the alloying reaction. (d) We track the average porosity of the electrode during the alloying reaction. Starting from an initial value of 33$\%$, the pore area decreases logarithmically with increasing SOC, reaching a final porosity of $\sim 14\%$. The peak stresses $\sigma_{yy}$ at the lower right pore is plotted as well. We note that the pores are mostly under compression (except at SOC = 11$\%$ and 65$\%$ that are related to sharp changes in pore geometry). We note that the critical stress value accompanying the Li$_2$Sb to Li$_3$Sb is $\sigma_c = 1.48$ GPa.}
    \label{fig: pores}
\end{figure}

\vspace{2mm}
\noindent Fig.~\ref{fig: pores} shows four ideal (no sharp corners, symmetric) pores arranged periodically in antimony with fixed displacement $\mathbf{u}=0$ on the particle boundary. The pores occupy $33\%$ of the electrode volume fraction (consistent with our experimental samples) and are large enough for the electrolyte to seep through the pores. On lithiating, Li$_3$Sb phase nucleates around the corners of the pores and grows rapidly to a large domain, see Fig.~\ref{fig: pores}(b). This growth of the two-phase microstructure is accompanied by $\sim24\%$ volume change that significantly deforms the pores. For example, as shown in Fig.~\ref{fig: pores}(d) (solid black line), the porosity of the electrode rapidly reduces to $22.29\%$ at SOC = $20.2\%$ and to $\sim 14\%$ at $100\%$ SOC.

\vspace{2mm}
\noindent Fig.~\ref{fig: pores} (bottom row) shows the stress distribution in a porous electrode during lithiation. At SOC = $10-20\%$, we note a peak in the tensile stresses across the phase boundary of $\sigma_{yy} = 2$ GPa, however, the stresses at the pore walls are negligible $\sigma_{yy} = -0.18$ GPa. With continued lithiation, the electrode swells into the pores changing their geometries (see pointed corners in Fig.~\ref{fig: pores}(c)). A majority of the pores are under compression $\sigma_{yy} = -10.1$ GPa, but these stresses evolve with the changing pore geometry. For example, Fig.~\ref{fig: pores}(d) shows the peak stresses around the lower right pore that are primarily under compression, however, there are two discontinuities at SOC $\approx 11\%$ and $65\%$. Compressive stresses indicate that the Li$_3$Sb phase swells into the pore, accommodating the large volume change, and the tensile peaks correspond to the sharp changes in the pore geometry. Despite these stresses at the pores, it is important to note that most of the stress in nanoporous electrodes is typically 2-3 times smaller than the interfacial stresses at crack tips or in the bulk electrode particle, see Figs.~\ref{fig: pores}, \ref{Fig2}, and \ref{fig: cracks}(a). These nanoporous features accommodate the volume changes and relieve internal stresses in alloying electrodes, and would explain the relatively longer lifespans of nanoarchitected electrodes in Section~\ref{sec:experiments}.

\begin{figure}[ht!]
    \centering
    \includegraphics[width=0.8\textwidth]{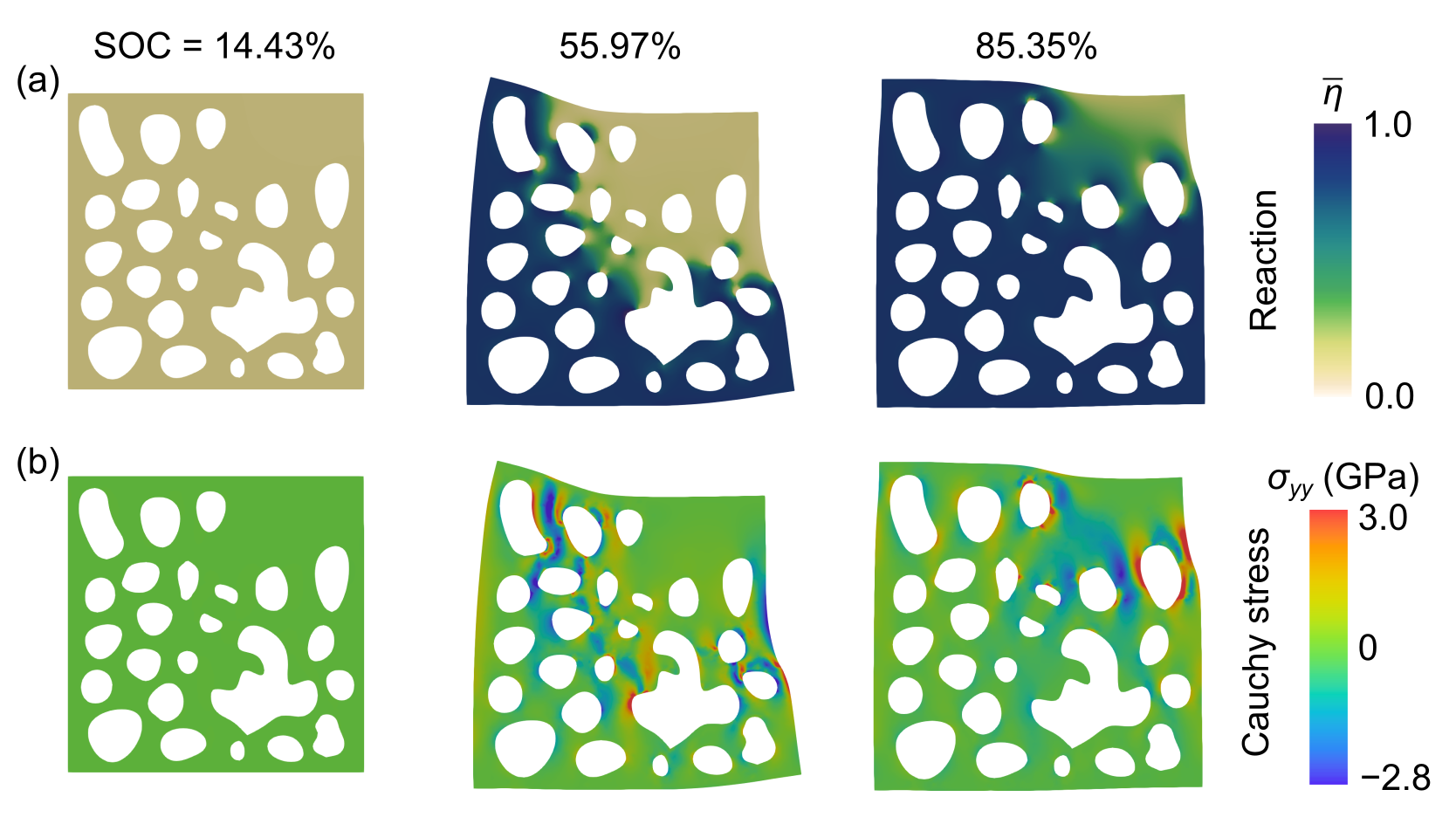}
    \caption{Nanoporous electrode geometries extracted from micrographs in Ref.~\cite{lin2020understanding} show reduced diffusion lengths and accommodate large volume changes during the alloying reaction. (a) The top row shows the nanopores accommodating volume changes during the lithiation cycle. (b) The corresponding Cauchy stresses are primarily concentrated around the phase boundary and at sharp corners of the nanopores.}
    \label{fig: experimental-pore-spinodal}
\end{figure}

\vspace{2mm}
\noindent Fig.~\ref{fig: experimental-pore-spinodal} shows stresses around interconnected pores during the alloying reaction. We binarize the experimental micrographs of pores in Sb (e.g., from Ref.~\cite{lin2020understanding} and Fig.~\ref{Exp_F1} in the experimental section~\ref{sec:experiments}), and develop an image extraction algorithm to model realistic pore geometries in our finite element calculations. During lithiation (at 15$\%$ SOC), we note that the Li$_3$Sb phase nucleates rapidly around the pores in comparison to the bulk region (see upper-right corner of the particle in Fig.~\ref{fig: experimental-pore-spinodal}). We attribute this to the increased exposure of the porous regions to the electrolyte and reduced diffusion lengths. The heterogeneous nucleation around pores quickly merges into a single reacted phase and the random distribution of pores disrupts the core-shell transformation pathways observed in defect-free particles. 
At SOC $\approx 56\%$, we note tensile stresses of $\sigma_{yy}$ = 3.0 GPa at the sharp edges of interconnected pores that are subsequently relieved with continued and affine volume accommodation into the pores. 

\subsection{Designing \textcolor{black}{Geometric Instabilities} in Electrodes}
\begin{figure}[ht!]
    \centering
    \includegraphics[width=\textwidth]{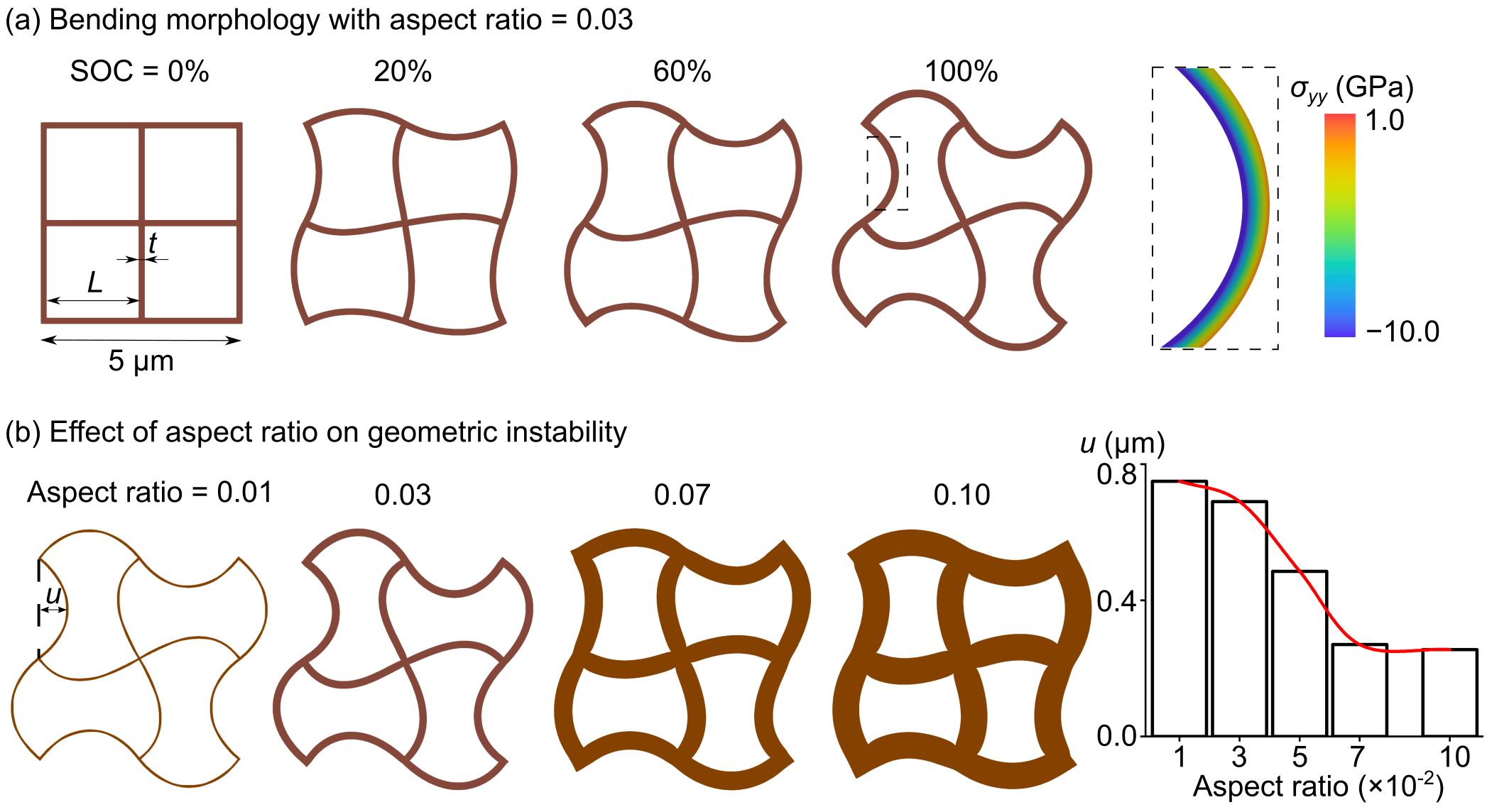}
    \caption{A demonstration of an architected alloying anode that accommodates finite deformations by bending systematically. (a) We model a periodic tessellation of square lattices (of side $L = 2.5\mu\mathrm{m}$ and a wall thickness $t$). For an electrode geometry with an aspect ratio of $t/L = 0.03$, lithiating the anode shows affine bending of the electrode sides.  For example, the walls are displaced inward, accommodating volume changes and alleviating internal stresses. A peak tensile stress $\sigma_{yy} < 1$~GPa is observed during lithiation, which is below the critical stress value for failure in Sb electrodes. (b) We vary the aspect ratio of the lattices $\frac{t}{L}$ from 0.01 to 0.1 and simulate the charging process for each configuration. We measure the maximum horizontal deflection $u$ as the displacement relative to the reference square lattice. Significant bending occurs for aspect ratios below 0.07, while this deformation is negligible at higher aspect ratios where the walls are too thick to undergo substantial bending. }
    \label{fig: geometric-instability}
\end{figure}

\noindent Our computations in Fig.~\ref{fig: geometric-instability} highlight an advantage of our continuum model as a potential design tool for architecting nanoscale features in alloying anodes. In this subsection, we explore one such electrode architecture, in which we engineer the bending of slender beams to reversibly accommodate the large volume changes accompanying alloying reactions. This design contributes to improved structural reversibility, however, compromises on the energy storage density. All factors must be considered in optimizing the electrode design. For our purposes, we explore the mechanics of these electrode architectures from the perspective of structural longevity, reversible (and repeatable) phase transformations, and potentially faster diffusion pathways.

\vspace{2mm}
\noindent We model an electrode architecture consisting of a tessellation of square lattices with varying aspect ratios $\frac{t}{L}$, as shown in Fig.~\ref{fig: geometric-instability} (inspired by designs in Refs.~\cite{liu2016harnessing, xia2016situ}). The reference square-structured electrode features particles with a size of 5 $\mu$m, consistent with the typical particle size in the experiments. Each particle comprises four small square lattices with side length $L$, separated by walls of thickness $t$. These dimensions satisfy the relationship $2L+3t=5\mu$m. We apply suitable periodic boundary conditions on the domain and Li-flux on all edges of the domain.

\vspace{2mm}
\noindent Fig.~\ref{fig: geometric-instability}(b) shows the deformation of architected electrodes for a range of aspect ratios $0.01 \leq \frac{t}{L} \leq 0.1$ when fully lithiated. We measure the maximum horizontal deflection $u$ as the displacement relative to the reference square lattice. For architectures with $\frac{t}{L} \leq 0.07$, we note that the nucleation and growth of the Li$_3$Sb phase induces non-uniform deformation in the electrode. In these cases, the electrode walls typically displace and bend to accommodate volume changes during alloying. For example, when the thickness-to-length ratio $\frac{t}{L}=0.01$, we observe the maximum deflection of approximately 0.73 $\mu$m, representing the largest bending deformation across all aspect ratios investigated in this study. By contrast, this bending behavior is less significant in electrode architectures with aspect ratios $\frac{t}{L} \geq 0.07$, in which the wall thickness is too thick to bend significantly.

\vspace{2mm}
\noindent Fig.~\ref{fig: geometric-instability}(a) shows the bending deformation in electrodes with an aspect ratio of 0.03. This bending can accommodate volume changes of up to $24\%$ with significantly lower internal stresses. Notably, our computation reveals that the peak stress in a fully lithiated square cell is approximately $\sigma_{yy} = 1.0~\mathrm{GPa}$ (see inset in Fig.~\ref{fig: geometric-instability}(a)), which is significantly lower than that observed in bulk particles and nanoporous structures. This stress level consistently remains below the threshold predicted by Griffith's criterion for antimony. Overall, these computations provide quantitative insights into how \textcolor{black}{geometric instabilities (i.e., bending)} induced by significant non-uniform volume changes can mitigate internal stresses and accommodate large volume changes.

\section{Discussion} \label{Discussion}
\noindent Using a combination of experiments and theory we investigate the finite deformations accompanying alloying reactions in anodes, and how the microstructural features interact with electrode geometries and collectively affect their structural reversibility. A key outcome of this work is a comprehensive micromechanical model, which takes into account the interplay between diffusion and reaction kinetics, to predict deformation mechanisms in nanoporous electrodes. For the first stage of lithiation (Sb $\to$ Li$_2$Sb), we model an elastic-plastic deformation and identify that the interfacial stresses are the primarily driver of the abrupt failure in bulk Sb electrodes. For the second stage of lithiation (Li$_2$Sb $\to$ Li$_3$Sb), our continuum model shows that the phase boundary interacts with defects, such as cracks and pores, and accumulates internal stresses during the alloying reaction. The theoretical framework is general and can be adapted to describe finite deformations in conversion or alloying electrodes.  The quantitative insights from simulations enable us to architect electrodes to improve structural reversibility and reduce diffusion pathways. Below we discuss key limitations of our model and then highlight the relevance of our results.

\vspace{2mm}
\noindent In our analytical study of the Sb $\to$ Li$_2$Sb transformation, we idealize the electrode to a 1D radially-symmetric core-shell geometry with uniform and isotropic material properties. This assumption does not account for irregular pore networks or localized stress concentrations that likely lead to microcracks in NP-Sb electrodes. Similarly, we use static material properties that do not evolve with the state of charge. These assumptions limit us from drawing general conclusions on failure in Sb electrodes; nevertheless, they emphasize the importance of porous geometries and mechanically soft phases in alleviating internal stresses.

\vspace{2mm}
\noindent In our continuum model, we use a neo-Hookean material law to predict the finite deformations accompanying the Li$_2$Sb to Li$_3$Sb alloying ($\approx 24\%$ volume change). In this reaction, both Li$_2$Sb and Li$_3$Sb are crystalline solids, and the neo-Hookean hyperelastic model appropriately captures the nonlinear stress-strain relationships evolving during the phase change. However, in other systems, such as sodiation of Sb \cite{selvaraj2020remarkable}, or the lithiation of Si \cite{zhao2011concurrent}, significantly larger volume changes---by an order of magnitude---occur, leading to plastic deformation of the electrode. In our future work, we plan to extend our theoretical framework for a Na/Sb system by incorporating elastic-plastic behavior, and investigate the interplay between microstructures and defects during an alloying reaction.

%\vspace{2mm}
%\noindent Alloying reaction in Li$_2$Sb is accompanied by an abrupt and large volume change ($\sim 24\%$) that generates a peak interfacial stress of 5 GPa during phase transformation. Our computations show that these stresses are further increased at sharp corners, such as crack tips, and exceed the critical stress value necessary for fracture in Sb electrodes. We note that these stresses can be reduced for a suitable combination of diffusion and reaction kinetics (i.e., diffusion kinetics are much faster than the reaction kinetics cases with $\mathrm{D_0} > \mathrm{R_0}$) and defect geometries such as a smooth pore relieve interfacial stresses. Furthermore, pore-like geometries deform during phase transformation and accommodate the large volume changes in Sb electrodes. For example, a nanoporous electrode (with $33\%$ volume fraction of pores) reduces the overall expansion of the electrode by $20\%$. Interconnected pores with sharp corners, however, accumulate internal stresses, which with repeated cycling could lead to particle fracture. These calculations compare favorably with our experimental observations of electrode fracture in bulk Sb ($\sim 5\mu\mathrm{m}$ particle size) and improved reversibility (40 cycles) in nanoporous Sb electrodes.

\vspace{2mm}
\noindent We demonstrate an application of our theoretical framework as a design tool and engineer an electrode architecture in which geometric instabilities such as bending are designed for improved structural reversibility. Specifically, we show that an electrode geometry comprising a periodic tessellation of square unit cells undergoes reversible transformations with small tensile stresses during alloying. That is, during lithiation the individual electrode members bend to accommodate the large volume changes and relieve the internal stresses, and return to their initial configuration on delithiation. This reversible deformation is also repeatable in that the microstructural evolution pathways are similar in each electrochemical cycle. This electrode design can be further optimized for efficient diffusion kinetics and energy density, in addition to the improved structural reversibility of the alloying electrodes.

\vspace{2mm}
\noindent Overall, we note that the continuum model describes crystalline-to-crystalline phase transformations accompanying an alloying reaction, and provides quantitative insights into the interplay between microstructural changes and deformation mechanisms in nanoarchitected electrodes (e.g., pores). In future work, this modeling framework can be generalized to predict the formation of intermediate disordered phases (e.g., in Na/Sb system) and how these phases facilitate stress relaxation. Numerical techniques such as the level-set methods \cite{van2020rechargeable,gibou2018review} would be useful to track phase boundary movement and describe the coherency at the interface between a crystalline/amorphous phase \cite{brady2024multiscale}. 

\section{Conclusion} \label{Conclusion}
\noindent Our work provides fundamental insights into the failure mechanisms associated with finite deformations in alloying anodes. Using a combination of micromechanical theory and electrochemical experiments, we demonstrate the hierarchical interplay between microstructural patterns and electrode geometries during phase changes in the Li/Sb system. Our experiments show a clear phase transition from Sb $\to$ Li${_2}$Sb $\to$ Li$_3$Sb and an improved structural reversibility in NP-Sb electrodes. In Sb $\to$ Li$_2$Sb, our analytical model predicts that the interfacial stresses at Sb/Li$_2$Sb are a primary factor contributing to the cracking of Sb electrodes. Additionally, our analysis shows that nanoporous geometries reduce hoop stresses when compared to the bulk, and therefore, suppress the propagation of microcracks $< 0.5~\mu\textrm{m}$. For Li$_2$Sb$\to$Li$_3$Sb, our variational model provides microstructural insights into the stresses and anisotropic volume changes accompanying the phase transformation. Specifically, we find that hollow cavities in nanoporous geometries act as buffers by accommodating large volume changes (up to $\sim 20\%$) and reducing peak stresses at the pore walls. These findings highlight the potential of engineering electrode geometries with pores (or lattice structures) to minimize internal stresses and improve structural reversibility.

\section*{Acknowledgments}
\noindent  Y.-C. Lai  and A. Renuka Balakrishna acknowledge support by the U.S. Department of Energy, Office of Science, Basic Energy Sciences, under Award DE-SC0024227 (variational methods development, numerical simulations). All authors gratefully acknowledge research funding by the U.S. Department of Energy’s (DOE) Energy Earthshot (StORE) under Award DE-FOA-0003003 (analytical model development: DZ, YCL, ARB; synthesis of samples and characterization: KT, KS, JNW, ST). Y.-C. Lai  and A. Renuka Balakrishna acknowledge the Center for Scientific Computing at the University of California, Santa Barbara (MRSEC; NSF DMR 2308708) for providing computational resources that contributed to the results reported in this paper. Use of the Stanford Synchrotron Radiation Lightsource, SLAC National Accelerator Laboratory, is supported by the U.S. Department of Energy, Office of Science, Office of Basic Energy Sciences under Contract No. DE-AC02-76SF00515. 

\newpage
\bibliographystyle{elsarticle-num} 
\bibliography{Sb}

\newpage
\setcounter{figure}{0}
\setcounter{equation}{0}
\begin{appendices}     
\appendix
\renewcommand{\thesection}{Appendix \Alph{section}}
\clearpage
\section{Symbols and Material Parameters} \label{A}
\noindent We summarize all symbols and material parameters used in our work in Tables~\ref{Table-Constants} and \ref{tab:material constants} below.
\begin{table}[ht!]
  \centering
  \caption{A brief description of variables introduced in our modeling framework}
\renewcommand\arraystretch{1.5}\begin{tabular}{lll}
  \hline
  Variable & Description \\
    \hline
    $\eta$ & Extent of alloying reaction \\
    $c$ & Lithium concentration \\    
    $\mathbf{N}$ & Direction of reaction  \\
    $\mathbf{F}$ & Deformation gradient \\
    $\mathbf{F}_0$ & Chemical component of $\mathbf{F}$\\
    $\bar{\mathbf{F}}$ & Elastic component of $\mathbf{F}$  \\
    $\bar{\mathbf{C}}$ & Cauchy-Green deformation tensor $\bar{\mathbf{C}}=\bar{\mathbf{F}}^{\top}\bar{\mathbf{F}}$\\
    $\mathrm{J}$ & Total volume change, $\mathrm{J}=|\mathrm{det}\mathbf{F}|$ \\
    % \textcolor{black}{$\Delta \mathrm{V_e}$} & Elastic volume \textcolor{black}{ratio}, $\Delta \mathrm{V_e}=|\mathrm{det}\mathbf{F}_\mathrm{e}|$  \\    
    $\mathrm{J}_0$ & Chemical volume change, $\mathrm{J}_0=|\mathrm{det}\mathbf{F_0}|$ \\ %$\mathrm{J}_0=|\mathrm{det}\bar{\mathbf{F}}|$
    $\boldsymbol{\sigma}$ & Cauchy stress tensor  \\
    $\mathbf{P}$ & First Piola-Kirchhoff stress tensor \\
    $\bar{\mathbf{S}}$ & Second Piola-Kirchhoff stress tensor \\
    $\bar{\mathbf{M}}$ & Mandel stress tensor \\
    \hline
    \end{tabular}
    \label{Table-Constants}
\end{table}

\newpage
\begin{table}[ht!]
    \centering
    \caption{List of material constants corresponding to the Li$_2$Sb-Li$_3$Sb system}
    \renewcommand\arraystretch{1.5}
    \begin{tabular}{lll}
    \hline
    Notation&Description&Value\\
    \hline
    $\mathrm{c_{max}}$ & Maximum concentration of diffusing species & 0.17 $\times$ 10$^{6}$ [mol/m$^{3}$]\\
    $\mathrm{\eta_{max}}$ & Maximum concentration of reacted species & 0.03 $\times$ 10$^{6}$ [mol/m$^{3}$]\\
    $E_a$ & Reaction energy barrier & $0.6\cdot 10^6$ [kJ/$\mathrm{m^3}$] \cite{chang2015elucidating}\\
    % $\mu_0^{\alpha}$ & Reference potential of species& 0 [kJ/mol]\\
    % &diffusing in the unreacted phase&\\
    $\mu_0^{\eta}-\mu_0^{\alpha}$ & Chemical potential difference Li$_2$Sb/Li$_3$Sb& $- 17.5$ [kJ/mol] \cite{terlicka2016enthalpy}\\
    % $\mathrm{R}$ & Gas constant & 8.314 [J/mol$\cdot$K] \\
    % $\mathrm{T}_0$ & Reference Temperature & 300 [K] \cite{chang2015elucidating} \\
    $G$ & Shear  modulus & 26 [GPa] \cite{jain2013commentary}\\
    $\lambda$ & First Lam\'e constant & 15.6 [GPa] \cite{jain2013commentary}\\
    % $\mathrm{R}_0$ & Reaction kinetic constant & 8 $\times$ 10$^{-6}$ [1/s]\\
    $\mathrm{D_0}$ & Diffusion constant & $4\times10^{-13}$ [m$^2$/s] \cite{allcorn2015lithium}\\
    $\Omega c^{\mathrm{R}}_0$ & Volume change from Li$_2$Sb to Li$_3$Sb& 1.24 \cite{chang2015elucidating}\\
    $\kappa$ & Gradient coefficient & 2 $\times$ 10$^{-9}$ [J/m] \cite{chen2014phase}\\
    \hline
    \end{tabular}
    \label{tab:material constants}
\end{table}

\begin{enumerate}
    \item The hexagonal Li$_2$Sb belongs to space group $P\bar{6}2c$ (no. 190).
    \item The reacted phase is Li$_3$Sb which has cubic symmetry and belongs to the space group $Fm\bar{3}m$ (no. 225).
    \item The maximum concentration of diffusing species is given by $\mathrm{c_{max}}=\phi\rho_{\mathrm{Li_2Sb}}/\mathcal{M}_{\mathrm{Li_2Sb}}$ with $\rho_{\mathrm{Li_2Sb}}=3.79~\mathrm{g/cm^{3}}$ the mass density of the host, $\mathcal{M}_{\mathrm{Li_2Sb}}=135.64~\mathrm{g/mol}$ the molar mass of the host, and $\phi$ the number of sites available. We use here $\phi=6$ due to the hexagonal-type structure of the Li$_2$Sb crystal.
    \item The maximum concentration of reacted species is given by $\mathrm{\eta_{max}}=1\rho_{\mathrm{Li_2Sb}}/\mathcal{M}_{\mathrm{Li_2Sb}}$, where the coefficient 1 is the stoichiometric amount of Li in the reacted product in the reaction.
\end{enumerate}

\clearpage
\section{Analytical Model for a Spherical Electrode} \label{B}
\setcounter{figure}{0}
\noindent In this section, we outline the analytical model used to derive interfacial stresses at the phase boundary in a partially lithiated antimony particle. Consistent with experimental observations, we model a spherical electrode of radius $a_0$=5~$\mathrm{\mu m}$ with bulk and nanoporous geometries. On lithiation, these electrode particles expand isotropically, and we set the initial reaction front position at $c_0$ = $\alpha a_0$ for some $0\leq\alpha\leq1$, see Figs.~\ref{fig:Model}(a-b). The lithiated phase Li$_2$Sb nucleates and grows on the surface of the particle, resulting in a reaction front layer separating the unreacted Sb core from the Li$_2$Sb shell.

\begin{figure}[!ht]
    \centering   
    \includegraphics[width=0.75\textwidth]{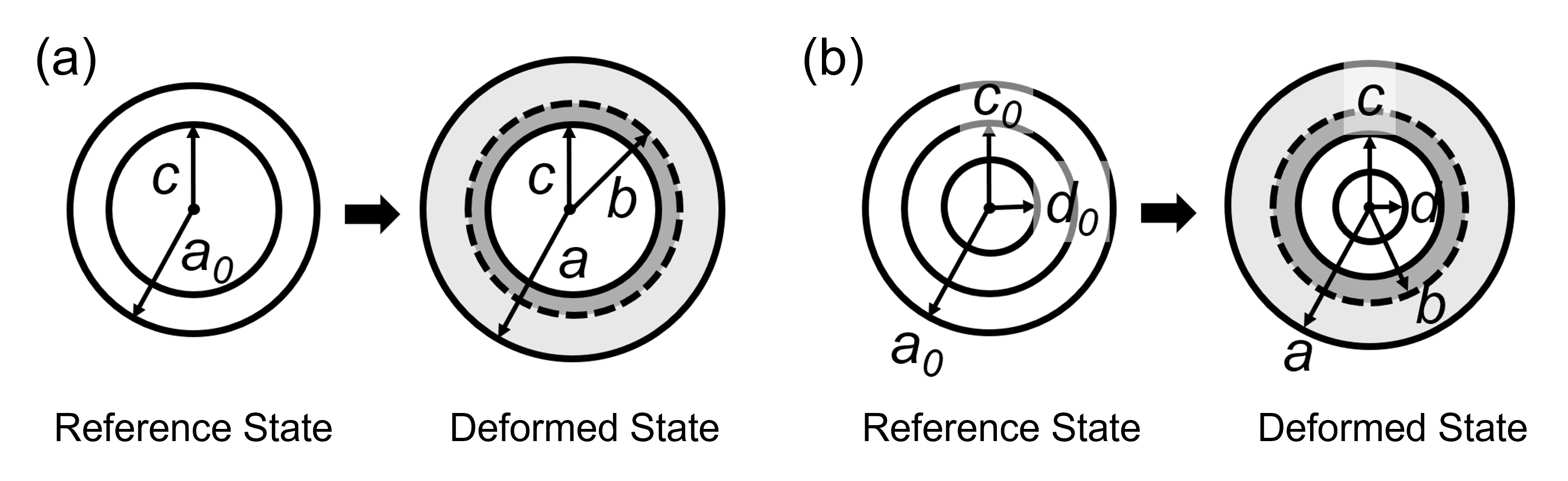}
    \caption{A schematic illustration of the different stages of lithiation (Sb $\rightarrow$ Li$_2$Sb) in (a) bulk and (b) nanoporous antimony particle.}
    \label{fig:Model}
\end{figure}

%\sout{For region between the sphere surface and the reaction front is the lithiated phase, and will undergo a volume expansion with rate $\beta$. Within this phase, the expansion is assumed to be large enough to enable the exsitence of plastic deformation region. Since the largest lattice misfit sits at the reaction front, we assume the plastic region to be somewhere between the surface and the reaction front obtained by ratio $\gamma$. As for the non-reacted region within reaction front, it is assumes to be rigid.}

\vspace{2mm}
\noindent In the bulk electrode, the outer shell expands radially and undergoes a plastic deformation (to accommodate the 90$\%$ volume change), and we assume that the core is rigid. The reaction front layer is a plastically deformed region sandwiched between the unlithiated core and the lithiated shell of Sb particle. Fig.~\ref{fig:Model}(a) shows a partially lithiated bulk particle (SOC = 50$\%$). The position of the deformed spherical surface and the reaction front are determined as follows:

\begin{align}
&a = \left[c^3 + \beta\left(a_0^3 - c^3\right)\right]^{1/3},\\
&b = c + (a-c)\gamma,\\
&c = \alpha a_0.
\end{align}

\noindent The coefficient $\beta$ = 1.9 corresponds to the volume ratio of the Li$_2$Sb to the Sb phases \cite{baggetto2013intrinsic}, and the coefficients $\gamma = 0.5$ and  $\alpha = 0.8$ represent the position of the reaction front layer.

\vspace{2mm}
\noindent Unlike the bulk electrode, the nanoporous electrode accommodates deformation through both outward expansion of the shell and inward shrinking of the pore. Fig.~\ref{fig:Model}(b) shows a $5\mu\textrm{m}$ sphere with a pore radius 
of $d_0=\delta a_0$. In our calculations, $\delta = 0.7$ that is comparable to the porosity measured in experimentally synthesized electrode particles. On lithiation, a Li$_2$Sb shell nucleates on the outer surface of the NP-Sb particle. The geometric positions of the particle surface, reaction-front layer, and the pore are determined as follows:

%During lithiation, the inner pore also participated in accommodating deformation through inward shrinking with rate $\eta$. The rest of the geometric value for deformed state can be determined by equilibrium of reacted/non-reacted region's volume:
\begin{align}
&a = \left[c^3 + \beta\left(a_0^3 - c_0^3\right)\right]^{1/3}\\
&b = c+(a-c)\gamma\\
&c = \left[d^3 + \left(c_0^3 - d_0^3\right)\right]^{1/3}\\
&d = \eta a_0,
\end{align}

\noindent with $\eta$ representing the change in pore radius. 

\vspace{2mm}
\noindent The solutions for the stress fields in Fig.~\ref{fig:lithiation} of the main text are derived by solving Eqs.~\ref{Force_equilibrium}--\ref{Eq: Plastic stress} subjected to the following boundary and continuity conditions:
\begin{itemize}
    \item For the bulk Sb particle (Fig.~\ref{fig:Model}(a)), we consider a traction-free boundary condition $\sigma_r(a) = 0$, stress continuity at the reaction front layer $\sigma_r(b^+)=\sigma_r(b^-)$, $\sigma_r(c^+)=\sigma_r(c^-)$, a yield criteria at the elastic-plastic interface $\sigma_\theta (b)-\sigma_r(b) = \sigma_Y $, and a hydrostatic compressive core $\sigma_r(r<c)=\sigma_\theta (r<c)$. 
    \item For the nanoporous model (Fig.~\ref{fig:Model}(b)), we consider traction-free boundaries $\sigma_r(a) = \sigma_r(d)= 0$, stress continuity across the reaction front layer $\sigma_r(b^+)=\sigma_r(b^-)$, $\sigma_r(c^+)=\sigma_r(c^-)$, a yield criteria at the elastic-plastic interface $\sigma_\theta (b)-\sigma_r(b) = \sigma_Y $. 
\end{itemize}

\clearpage
\section{Numerics for the Variational Model} \label{C}
\noindent We present the finite element implementation of a diffusion-reaction process incorporating finite deformation theory for alloying reactions. This represents a coupled, nonlinear, initial boundary value problem in which the Cahn-Hilliard type of diffusion, together with linear reaction kinetics, and mechanical equilibrium formulation are solved.

\vspace{2mm}
\noindent Recall that the local mass balance in Eqs.~(\ref{eq:reaction kinetics})-(\ref{eq:reaction chemical potential}) is
\begin{align}
    \frac{\partial c}{\partial t} &= \nabla\cdot[\mathbf{M}(c)\nabla\mu] - \frac{\partial \eta}{\partial t},\label{eq:12}\\
    \nonumber\\
    \mu&=\frac{\partial \psi}{\partial c}-\underbrace{\nabla \cdot \frac{\partial \psi}{\partial \nabla c}}_\text{=0},\label{eq:13}\\
    \nonumber\\
    \frac{\partial \eta}{\partial t} &= -\frac{\mathrm{R_0}\eta_{\mathrm{max}}}{\mathrm{RT}}[\frac{\partial \psi}{\partial \eta} - \frac{\Omega\bar{\eta}}{3}\mathrm{tr}(\bar{\mathbf{M}})-\nabla \cdot \frac{\kappa}{\eta_{\mathrm{max}}}\nabla \bar{\eta}-\frac{\partial \psi}{\partial c}].\label{eq:14}
\end{align}

\vspace{2mm}
\noindent The Galerkin weak form of the mixed formulation is obtained by multiplying Eqs.~(\ref{eq:12})-(\ref{eq:14}) with suitable test functions $\delta \mu$, $\delta c$, and $\delta \bar{\eta}$, respectively, and integrating over the domain $\Omega$. Starting with Eq.~(\ref{eq:12}), we have:
\begin{align}
0= \int_{\Omega}(\frac{\partial c}{\partial t}+\frac{\partial \eta}{\partial t})\delta \mu~dV+  \int_{\Omega} \mathbf{M}(c) \nabla \mu \cdot \nabla(\delta \mu)~dV-\int_{\partial \Omega}(\mathbf{M}(c)\nabla \mu  \cdot \hat{\mathbf{n}} )\delta \mu ~dA.
\label{eq:B4}
\end{align}

\vspace{2mm}
\noindent Similarly, for Eq.~(\ref{eq:13}), we have: 
\begin{align}
0= \int_{\Omega}\frac{\partial \psi}{\partial c}\delta c~dV- \int_{\Omega}\mu\delta c~dV.
\label{eq:B5}
\end{align}

\vspace{2mm}
\noindent For Eq.~(\ref{eq:14}), we have: 
\begin{align}
0&=\int_{\Omega}\frac{\partial \bar{\eta}}{\partial t} \delta \bar{\eta}~dV+\int_{\Omega}\frac{\mathrm{R_0}}{\mathrm{RT}}[\frac{\partial \psi}{\partial \eta} - \frac{\Omega\bar{\eta}}{3}\mathrm{tr}(\bar{\mathbf{M}})-\frac{\partial \psi}{\partial c}]\delta\bar{\eta}~dV\nonumber\\
&+\int_{\Omega}\frac{\mathrm{R_0}}{\mathrm{RT}}\frac{\kappa}{\eta_{\mathrm{max}}}\nabla \bar{\eta}\cdot \nabla(\delta\bar{\eta})~dV
-\int_{\partial \Omega}\frac{\mathrm{R_0}}{\mathrm{RT}}\frac{\kappa}{\eta_{\mathrm{max}}}(\nabla \bar{\eta}\cdot\hat{\mathbf{n}})\delta\bar{\eta}~dA.
\label{eq:B6}
\end{align}

\vspace{2mm}
\noindent Recall that the local force balance on Eq.~(\ref{eq:force balance}) is
\begin{align}
    \nabla\cdot\mathbf{P}^{\top} = 0.
    \label{eq:B7}
\end{align}

\vspace{2mm}
\noindent We multiply Eq.~(\ref{eq:B7}) with variational test function $\delta \mathbf{u}$, and the weak form is
\begin{align}
0=\int_{\Omega}\mathbf{P}:\nabla\delta\mathbf{u}~dV-\int_{\partial\Omega}(\mathbf{P}\cdot\hat{\mathbf{n}})\cdot\delta\mathbf{u}~dA.
\label{eq:B8}
\end{align}

\vspace{2mm}
\noindent Finally, we implement the above weak forms Eq.~(\ref{eq:B4}), (\ref{eq:B5}), (\ref{eq:B6}) and (\ref{eq:B8}) in the open source finite-element, multiphysics framework MOOSE \cite{gaston2009moose}. We solve the system of nonlinear equations using the preconditioned Jacobian-free Newton-Krylov (PJFNK) method. This approach does not require defining an explicit tangent matrix and, therefore, saves considerable computational time and storage. We further eliminate the rigid body modes at the solver level to prevent arbitrary rigid body displacements and rotations. This approach demonstrates improved convergence compared to conventional methods that constrain displacements at specific vertices. We use the implicit Backward-Euler method for time integration and an adaptive time-stepping approach for the relatively smooth diffusion process.

\end{appendices}

\end{document}